\newif\ifconfver
\confverfalse      

\newif\ifcutshort  
\cutshorttrue
\newif\ifcutshortlvltwo  
\cutshortlvltwofalse

\ifconfver
    \documentclass[10pt,twocolumn,twoside]{IEEEtran}
\else
    \documentclass[11pt,draftcls,onecolumn,twoside]{IEEEtran}
\fi
\usepackage{cite}
\usepackage{graphicx}
\usepackage{psfrag}
\usepackage{subfigure}
\usepackage{url}
\usepackage{stfloats}
\usepackage{amsfonts,amssymb,amsmath,bm,paralist,theorem,cite,ifthen,color}
\usepackage{array}
\usepackage{calc}
\usepackage{subfig}
\usepackage{longtable}
\usepackage{stfloats}
\usepackage{multirow}
\usepackage{algorithmic,algorithm}
\usepackage{graphics,booktabs,epsfig,enumerate}
\usepackage{amsfonts,amssymb,amsmath,amsbsy,bm,paralist,theorem,cite,ifthen,color}

{\begin{list}{}{
    \settowidth{\labelwidth}{\mbox{\textnormal{#1}}}%
    \setlength{\leftmargin}{\labelwidth+\labelsep}}}%
{\end{list}}

\newcommand{\Rset}  {\ensuremath{\mathbb{R}}}
\newcommand{\Cset}  {\ensuremath{\mathbb{C}}}

\newcommand{\Kset}  {\ensuremath{\mathcal{K}}}
\newcommand{\Lset}  {\ensuremath{\mathcal{L}}}
\newcommand{\Nset}  {\ensuremath{\mathcal{N}}_c}

\newcommand{\bb}        {\ensuremath{\mathbf{b}}}
\newcommand{\eb}        {\ensuremath{\mathbf{e}}}
\newcommand{\fb}        {\ensuremath{\mathbf{f}}}
\newcommand{\gb}        {\ensuremath{\mathbf{g}}}
\newcommand{\hb}        {\ensuremath{\mathbf{h}}}

\newcommand{\tb}        {\ensuremath{\mathbf{t}}}
\newcommand{\vb}        {\ensuremath{\mathbf{v}}}
\newcommand{\wb}        {\ensuremath{\mathbf{w}}}
\newcommand{\xb}        {\ensuremath{\mathbf{x}}}
\newcommand{\yb}        {\ensuremath{\mathbf{y}}}
\newcommand{\zb}        {\ensuremath{\mathbf{z}}}

\newcommand{\Ab}        {\ensuremath{\mathbf{A}}}
\newcommand{\ab}        {\ensuremath{\mathbf{a}}}

\newcommand{\Eb}        {\ensuremath{\mathbf{E}}}
\newcommand{\Fb}        {\ensuremath{\mathbf{F}}}
\newcommand{\Ib}        {\ensuremath{\mathbf{I}}}
\newcommand{\Qb}        {\ensuremath{\mathbf{Q}}}

\newcommand{\Wb}        {\ensuremath{\mathbf{W}}}
\newcommand{\Ub}        {\ensuremath{\mathbf{U}}}
\newcommand{\Xb}        {\ensuremath{\mathbf{X}}}
\newcommand{\Yb}        {\ensuremath{\mathbf{Y}}}
\newcommand{\Zb}        {\ensuremath{\mathbf{Z}}}

\newcommand{\xib}       {\ensuremath{{\bm{\xi}}}}
\newcommand{\nub}       {\ensuremath{{\bm{\nu}}}}

\newcommand{\Phib}      {\ensuremath{{\bm{\Phi}}}}
\newcommand{\Psib}      {\ensuremath{{\bm{\Psi}}}}

\newcommand{\hhb}       {\ensuremath{\hat{\hb}}}
\newcommand{\ghb}       {\ensuremath{\hat{\gb}}}

\newcommand{\zerob}     {\ensuremath{\mathbf{0}}}
\newcommand{\SINR}      {\ensuremath{{\rm{SINR}}}}
\newcommand{\Rank}      {\ensuremath{{\rm{Rank}}}}
\newcommand{\st}        {\ensuremath{{\rm{s.t.~}}}}
\newcommand{\MSnk}      {{\textrm{MS}}\ensuremath{_{nk}}}

\newcommand{\BSn}       {{\textrm{BS}}\ensuremath{_{n}}}
\newcommand{\BSm}       {{\textrm{BS}}\ensuremath{_{m}}}
\newcommand{\Tr}        {\ensuremath{{\mathrm{Tr}}}}
\newcommand{\cC}        {\ensuremath{{\mathcal{C}}}}

\newtheorem{Lemma}  {Lemma}
\newtheorem{Remark} {Remark}
\newtheorem{Prop}   {Proposition}

\begin{document}
\bibliographystyle{IEEEtran}

\title{Distributed Robust Multi-Cell Coordinated Beamforming with Imperfect CSI: \\An ADMM Approach$^\S$}

\ifconfver \else {\linespread{1.1} \rm \fi

\author{\vspace{0.5cm}Chao Shen$^\star$, Tsung-Hui Chang$^\dag$, Kun-Yu Wang$^\ddag$, Zhengding Qiu$^\star$, and Chong-Yung Chi$^\ddag$
\thanks{$^\S$
The work of Tsung-Hui Chang, Kun-Yu Wang and Chong-Yung Chi is supported by the National Science Council, R.O.C., under Grant NSC-99-2221-E-007-052-MY3.
The work of Chao Shen and Zhengding Qiu is supported by the National Basic Research Program of China (973 program) under Grant No. 2007CB307106.
Chao Shen is also supported by the Nufront Fellowship.
Part of this work was presented at the IEEE ICC, Kyoto, Japan, June 5-9, 2011 \cite{Shen2011}.}
\thanks{$^\dag$
Tsung-Hui Chang is the corresponding author.
Address: Institute of Communications Engineering, National Tsing Hua University, Hsinchu, Taiwan 30013, R.O.C.
E-mail: tsunghui.chang@gmail.com.}
\thanks{$^\star$
Chao Shen and Zhengding Qiu are with Institute of Information Science, Beijing Jiaotong University, Beijing, China, 100044.
E-mail: wwellday@gmail.com, zdqiu@bjtu.edu.cn.}
\thanks{$^\ddag$
Kun-Yu Wang and Chong-Yung Chi are with Institute of Communications Engineering \& Department of Electrical Engineering, National Tsing Hua University, Hsinchu, Taiwan 30013, R.O.C.
E-mail: kunyuwang7@gmail.com,~cychi@ee.nthu.edu.tw.}
}

\maketitle

\begin{abstract}
Multi-cell coordinated beamforming (MCBF), where multiple base stations (BSs) collaborate with each other in the beamforming design
for mitigating the inter-cell interference, has been a subject drawing great attention recently. Most MCBF designs assume perfect channel state information (CSI) of mobile stations (MSs); however CSI errors are inevitable at the BSs in practice. Assuming elliptically bounded CSI errors, this paper studies the robust MCBF design problem that minimizes the weighted sum power of BSs subject to worst-case signal-to-interference-plus-noise ratio (SINR) constraints on the MSs. Our goal is to devise a distributed optimization method that can obtain the worst-case robust beamforming solutions in a decentralized fashion, with only local CSI used at each BS and little backhaul signaling for message exchange between BSs. However, the considered problem is difficult to handle even in the centralized form.
We first propose an efficient approximation method in the centralized form, based on the semidefinite relaxation (SDR) technique.
To obtain the robust beamforming solution in a decentralized fashion, we further propose a distributed robust MCBF algorithm, using a distributed convex optimization technique known as \emph{alternating direction method of multipliers}
(ADMM). We analytically show the convergence of the proposed distributed robust MCBF algorithm to the optimal centralized solution and its better bandwidth efficiency in backhaul signaling over the existing dual decomposition based algorithms.
Simulation results are presented to examine the effectiveness of the proposed SDR method and the distributed robust MCBF algorithm.
\ifconfver
\else
\\\\
\noindent {\bfseries EDICS}:  SAM-BEAM, MSP-APPL, MSP-CODR, SPC-APPL  \fi
\end{abstract}

\ifconfver \else \IEEEpeerreviewmaketitle} \fi

\ifconfver \else
\newpage
\fi

\section{Introduction}

Recently, multi-cell processing, or known as coordinated multi-point (CoMP), has drawn great attention because it can provide significant system throughput gains compared to the conventional single-cell designs \cite{Gesbert10JSAC,Irmer11COMM}. 
We consider the scenario where the base stations (BSs) are equipped with multiple antennas and the mobile stations (MSs) are equipped with single antenna. Each of the BSs employs transmit beamforming to communicate with the MSs within its cell. The BSs in different cells, according to the principle of interference coordination  \cite{Gesbert10JSAC}, collaborate with each other to jointly design the beam patterns in order to effectively mitigate the inter-cell interference (ICI). To this end, various multi-cell coordinated beamforming (MCBF) designs have been proposed \cite{Dahrouj10TWC,Venturino10TWC,Bjornson2010}.
Most of the MCBF designs assume that the BSs are connected with a control center which knows all the MSs' channel state information (CSI) and computes the beamforming solution in a centralized manner. In practical multi-cell systems, however, obtaining the MCBF solutions in a decentralized fashion using only local CSI at each BS is of {central importance, thereby drawing} the developments of distributed beamforming design methods \cite{Dahrouj10TWC,Venturino10TWC,Bjornson2010,Nguyen2011TSP,Tolli11TWC,Zhangrui2010TSP,HuangTSP2011}.
The reasons are that 1) the future wireless systems prefer a flat IP architecture where all BSs are directly connected with the core network \cite{3GPP_standard}; 2) if the control center is {still employed, a distributed optimization method can be used to} decouple the original problem into multiple parallel subproblems with smaller problem size, thus reducing the required computation power of the control center \cite{Qiu2011TSP}.
By exploiting the uplink-downlink duality \cite{Farrokhi1998}, a distributed optimization method was proposed in \cite{Dahrouj10TWC}
for a power-minimization based MCBF design problem.
Game theory based distributed optimization methods were proposed recently in \cite{Nguyen2011TSP} for handling the same problem. In \cite{Tolli11TWC}, the dual decomposition technique \cite{Boyddecomposition} was used for developing a distributed optimization method {for the problem in \cite{Dahrouj10TWC}}. In \cite{HuangTSP2011}, the idea of uplink-downlink duality was used for distributed optimization of a different max-min-fair MCBF design problem.

The efficacy of beamforming designs relies on the assumption that the BSs have the perfect CSI of MSs. In practical scenarios, however, the BSs can never have perfect CSI, due to, e.g., imperfect channel estimation and finite rate feedback \cite{Love08JSAC}. In the multi-cell scenario, it is even more difficult for the BSs to obain reliable inter-cell CSI (i.e., the CSI of MSs that belong to the neighboring cells).
In the presence of CSI errors, the beamforming designs that {assume} perfect CSI will suffer from performance degradation, and the MSs' quality-of-service (QoS) requirements can no longer be guaranteed. In view of this, robust MCBF designs, which take into account the CSI errors, are of great importance. Assuming bounded CSI errors, e.g., quantization errors, a robust MCBF design was presented in \cite{Bjornson11TSP} which optimizes the worst-case symbol minimum mean squared error (MMSE) performance of the MSs.

{The focus of this paper is on} the worst-case signal-to-interference-plus-noise ratio (SINR) constrained MCBF design problem \cite{Tajer11TSP,Shen2011}, where the weighted sum power of BSs is minimized subject to constraints that guarantee worst-case SINR requirements for the MSs. Our goal is to develop a distributed beamforming optimization algorithm for the worst-case robust formulation; however, the considered problem itself is difficult to handle even in the centralized form, due to the fact that each of the worst-case SINR constraints in essence corresponds to infinitely many nonconvex constraints. The worst-case robust design formulation has been studied in the context of single-cell robust beamforming; see \cite{Shenouda07JSTSP,ZhengWongNg_2008}. However, the robust MCBF formulation is much more challenging since the associated worst-case SINR constraints involve CSI errors not only in the desired signal and intra-cell interference terms, but also in the inter-cell interference {term}.
To handle this problem, a convex restrictive approximation formulation is proposed in \cite{Tajer11TSP} which can provide feasible approximate solutions to the robust MCBF problem. Distributed optimization algorithms based on dual decomposition and {alternating} optimization are also presented in \cite{Tajer11TSP}. However, due to the reduced feasible set, the approximation formulation in \cite{Tajer11TSP} is less power efficient than the original problem.

In this paper, we propose a new convex approximation method for the worst-case SINR constrained robust MCBF design problem. Our approach is based on a convex approximation technique known as semidefinite relaxation (SDR) \cite{Luo2010_SPM}. 
SDR has been used in various transmit beamforming designs; see, e.g., \cite{BK:LuoChang}.
By SDR, we obtain a convex approximation formulation for the worst-case robust design problem, which, nevertheless, still involves complicated worst-case constraints. We decompose each of the worst-case constraints into several simpler worst-case constraints that can be conveniently handled by the S-lemma \cite{BK:BoydV04}. In particular, by the S-lemma, each decomposed worst-case constraint can be reformulated as a linear matrix inequality (LMI) \cite{ZhengWongNg_2008}. The resultant approximation formulation, which is a convex semidefinite program (SDP), thus can be efficiently solved by interior-point methods \cite{BK:BoydV04}. We also {identify several} conditions under which the proposed SDR method can yield the global optimal solution of the original worst-case robust MCBF problem.

We further {develop} a distributed optimization algorithm {that can solve the proposed SDR approximation formulation} in a decentralized fashion. While the dual decomposition method used in \cite{Tolli11TWC,Tajer11TSP} is conceptually applicable, we found that the resultant decomposed problems are numerically unstable due to lack of strict convexity.
To overcome this problem, we instead consider the so called {\textit{alternating direction method of multipliers}} (ADMM) \cite{BertsekasADMM,BoydADMM}. ADMM is an advanced dual decomposition method that combines the idea of dual decomposition and the augmented Lagrangian method \cite{BK:Bertsekas06}, where the latter is often used {for bringing numerical robustness to the dual accent method \cite{BoydADMM} by adding penalty terms for strict convexity of the problem}. Therefore, in contrast to the conventional dual decomposition method \cite{Boyddecomposition}, 
ADMM is more numerically stable and faster in convergence \cite{BertsekasADMM}. 
Based on the principle of ADMM, we propose a distributed robust MCBF algorithm that is provably able to converge to the global optimum of the centralized problem. In particular, by introducing some slack variables that represent the worst-case ICI powers, we decompose the SDR problem in a way such that the amount of messages required to be exchanged between BSs is much smaller than by the existing algorithms in \cite{Tolli11TWC,Tajer11TSP}, thus reducing {the bandwidth overhead of backhaul signaling}.

The rest of this paper is organized as follows. Section \ref{sec:signalmodel} presents the multi-cell system signal model and illustrates the impact of imperfect CSI on the performance of MCBF. The considered worst-case SINR constrained robust MCBF design formulation is also presented in that section. In Section \ref{sec: SDR}, the proposed SDR approximation method and its optimality conditions are presented. Using ADMM, the proposed distributed robust MCBF algorithm is presented in Section \ref{Sec:RobustDMCBF}.
Section \ref{sec:edge user} extends the proposed method to a fully coordinated scenario where some cell-edge MSs are served simultaneously by multiple BSs. Simulation results that demonstrate the effectiveness of the proposed {SDR method and distributed robust MCBF algorithm} are presented in Section \ref{Sec:Simulation}. Finally, conclusions are drawn in Section \ref{sec:conclusions}

%
\emph{Notations:}
$\Cset^n$, $\mathbb{R}^n$ and $\mathbb{H}^{n}$ stand for the sets of $n$-dimensional complex and real vectors and complex Hermitian matrices, respectively. $\mathbb{R}^n_+$ denotes the set of $n$-dimensional nonnegative orthant. Column vectors and matrices are written in boldfaced lowercase and uppercase letters, e.g., $\ab$ and $\Ab$. $\Ib_n$ denotes the $n\times n$ identity matrix, and $\zerob$ denotes an all-zero vector (matrix) with appropriate {dimension}. The superscripts $(\cdot)^T$, $(\cdot)^H$ and $(\cdot)^\dag$ represent the transpose, (Hermitian) conjugate transpose and pseudo inverse operations, respectively. $\Rank(\Ab)$ and $\Tr(\Ab)$ represent the rank and trace of matrix $\Ab$, respectively. $\Ab\succeq \zerob$ ($\succ \zerob$) means that matrix $\Ab$ is positive semidefinite (positive definite). For vector $\ab$, $\|\ab\|$ denotes the Euclidean norm. ${\mathbb E}\{\cdot\}$ denotes the statistical expectation. For a variable $a_{nmk}$, where $n\in \{1,\ldots,N\}$, $m\in \{1, \ldots,M\}$ and $k\in \{1,\ldots,K\}$, we denote $\{a_{nmk}\}_k$ as the set containing $a_{nm1},\ldots,a_{nmK}$; while we denote $\{a_{nmk}\}$ as the set containing all possible $a_{nmk}$, i.e., $a_{111},\ldots,a_{11K},a_{121},\ldots,a_{NMK}$.

\section{Signal model and Problem Statement}\label{sec:signalmodel}
This section presents the multi-cell downlink system model and the worst-case robust MCBF design problem. \vspace{-0.3cm}

\subsection{System Signal Model}\label{subsec: signal model}

Consider a multi-cell downlink system that consists of $N_c$ cells. Each cell is composed of one BS, which is equipped with $N_t$ antennas, and $K$ single-antenna MSs. The $N_c$ BSs are assumed to operate over a common frequency band and communicate with their $K$ respective MSs using transmit beamforming. The scenario under consideration is that each MS is served by only one BS; extension to the scenario where one MS is served by multiple BSs will be discussed in Section \ref{sec:edge user}.

We denote $\BSn$ as the $n$th BS, and $\MSnk$ as the $k$th MS in the $n$th cell, for all $n\in\Nset\triangleq\{1,2,\ldots,N_c\}$ and
$k\in\Kset\triangleq\{1,2,.\ldots,K\}$. Let $s_{nk}(t)\in\Cset$ be the information data stream for $\MSnk$, and $\wb_{nk}\in\Cset^{N_t}$ be the associated beamforming vector. The transmit signal by $\BSn$ is given by
\begin{align}\label{eq. transmit signal}
  \xb_n(t)=\sum_{k=1}^K{\wb_{nk}s_{nk}(t)},
\end{align}
for $n=1,\ldots,N_c$. The received signal of $\MSnk$ can be expressed as
\begin{align}\label{eq. received signal}
y_{nk}(t)
=&\sum_{m=1}^{N_c}\hb_{mnk}^H\left(\sum_{i=1}^K{\wb_{mi}s_{mi}(t)}\right)+z_{nk}(t) \notag\\ 
=&\hb_{nnk}^H\wb_{nk}s_{nk}(t)
 +\sum_{\substack{i\neq k}}^K{\hb_{nnk}^H\wb_{ni}s_{ni}(t)}                  
 +\sum_{\substack{m\neq n}}^{N_c}{\sum_{i=1}^K\hb_{mnk}^H\wb_{mi}s_{mi}(t)}  
 +z_{nk}(t),
\end{align}
where $\hb_{mnk}\in \Cset^{N_t}$ denotes the channel vector from $\BSm$ to $\MSnk$, and $z_{nk}(t)\in \Cset$ is the additive noise of $\MSnk$, which is assumed to have zero mean and {variance} $\sigma_{nk}^2>0$. The term $z_{nk}(t)$ may capture {both the receiver noise and} the interference from the other non-coordinated BSs. In \eqref{eq. received signal}, the first term is the signal of interest for $\MSnk$, and the second and third terms are the intra-cell interference and ICI, respectively.
Assume that $s_{nk}(t)$ are statistically independent, with zero mean and $\mathbb{E}\{|s_{nk}(t)|^2\}=1$ for all $n \in\Nset$ and $k \in \Kset$, and assume that each MS employs single-user detection. By \eqref{eq. received signal}, the SINR of $\MSnk$ is given by
\begin{align}\label{Eq:SINR1}
&\SINR_{nk}\left(\{\wb_{m1},\ldots,\wb_{mK}\}_{m=1}^{N_c},\{\hb_{mnk}\}_{m=1}^{N_c}\right)
=\frac{{\left|\hb_{nnk}^H\wb_{nk}\right|}^2}%
{\sum\limits_{i\neq k}^K {\left|\hb_{nnk}^H\wb_{ni}\right|}^2%
+\sum\limits_{m\neq n}^{N_c}\sum\limits_{i=1}^K {\left|\hb_{mnk}^H\wb_{mi}\right|}^2%
+\sigma_{nk}^2}.%
\end{align}
%
%

Conventional single-cell beamforming designs \cite{Gershman2010_SPM} are developed mainly for handling the intra-cell interference only. To take into consideration the ICI, the following multi-cell coordinated beamforming (MCBF) design has been considered \cite{Dahrouj2010}
\begin{subequations}\label{PM}
\begin{align}
  \min_{\substack{\wb_{nk}, k=1,\ldots,K\\ n=1,\ldots,N_c}}~
  &\sum_{n=1}^{N_c} {\alpha_n}\left({\sum_{k=1}^K {\left\|\wb_{nk}\right\|^2} }\right)\\
  {\rm \st}~~
  &\SINR_{nk}\left(\{\wb_{m1},\ldots,\wb_{mK}\}_{m=1}^{N_c},\{\hb_{mnk}\}_{m=1}^{N_c}\right)\geq \gamma_{nk} ~\forall
  k\in\Kset,~n\in\Nset,
\end{align}
\end{subequations}
where $\alpha_n>0$ is the power priority weight for $\BSn$.
As seen, the MCBF design jointly optimizes the beamforming vectors of all BSs such that the weighted sum power is minimized while {the MSs' SINR requirements $\gamma_{nk}>0$ can be fulfilled}. It has been shown that problem \eqref{PM} can be reformulated as a convex second-order cone program (SOCP) \cite{Dahrouj2010}, which can be efficiently solved via standard convex solvers, e.g., \texttt{SeDuMi} \cite{SeDuMi}.

\subsection{Worst-Case Robust MCBF Design}\label{subsec: worst case robust design}

\begin{figure}[t!]\centering%
  {\psfrag{figa}[c][c][0.9]{(a) Achievable SINR (dB): Non-robust MCBF \eqref{PM}}
   \psfrag{figb}[c][c][0.9]{(b) Achievable SINR (dB): Robust MCBF \eqref{RobustMCBF}}
  \includegraphics[width=0.47\linewidth]{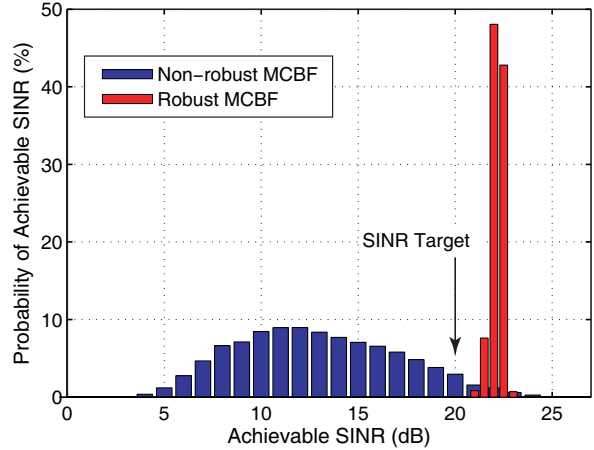}}%
  \caption{Distribution of the achievable SINR values
  of MS$_{11}$, when the non-robust MCBF design \eqref{PM} and the robust MCBF design
  \eqref{RPM} are respectively used.
  The target SINR value is 20 dB. %
}  \label{Fig:SINRhist}\vspace{-0.6cm}
\end{figure}

The MCBF design in \eqref{PM} assumes that the BSs have perfect knowledge of the CSIs. As discussed in the introduction, in practice, the BSs inevitably suffer from CSI errors.
Let $\hhb_{mnk} \in \mathbb{C}^{N_t}$, $n,m\in\Nset$, $k\in\Kset$ be the preassumed CSI at the BSs. Then the true CSI
can be expressed as
\begin{align}\label{CSIadditiveError}
\hb_{mnk}&=\hhb_{mnk}+\eb_{mnk}~\forall~ m,n\in\Nset,~k\in\Kset,
\end{align} where $\eb_{mnk}\in\Cset^{N_t}$
denotes the CSI error vector associated with the true channel $\hb_{mnk}$. Our interest lies in the bounded CSI errors. Specifically, we assume that each CSI error vector $\eb_{mnk}$ satisfies the following elliptic model:
\begin{align}\label{CSIerrormodel}
  \eb_{mnk}^H\Qb_{mnk}\eb_{mnk}\leq 1,
\end{align} where $\Qb_{mnk}\in \mathbb{H}^{N_t}$, $\Qb_{mnk} \succ \zerob$ specifies the size and
shape of the ellipsoid. When
$\Qb_{mnk}=(1/\varepsilon_{mnk}^2)\Ib_{N_t}$ where $\varepsilon_{mnk}^2>0$, \eqref{CSIerrormodel} reduces to the popular spherical error model $\|\eb_{mnk}\|^2 \leq\varepsilon_{mnk}^2$ \cite{Shenouda07JSTSP}.
In the presence of CSI errors, the non-robust design \eqref{PM} cannot guarantee the SINR requirement of MSs and consequently outage may occur. The following simulation example motivates the need of robust designs:

{\bf Example:} We consider a two-cell system ($N_c=2$), with two MSs in each cell ($K=2$). Each of the BSs is equipped with four antennas ($N_t=4$). A set of preassumed CSI $\{\hhb_{mnk}\}$ is randomly generated following the independent and identically distributed (i.i.d.) complex Gaussian distribution with zero mean and unit variance (see Section \ref{subsec:Simulation setting} for the detailed channel model used in the {simulation). Using} the preassumed CSI, an optimal beamforming solution is obtained by solving the MCBF problem \eqref{PM}, with a 20 dB target SINR for all the four MSs (i.e., $\gamma_{nk}=20$ dB for $n,k=1,2$). To examine the impact of the CSI errors, we randomly generate 10,000 sets of CSI errors satisfying the following spherical error model
\begin{align*}
   &\|\eb_{nnk}\|^2 \leq 0.01,~\|\eb_{mnk}\|^2 \leq 0.04~\forall~m\neq n,
\end{align*} for $n=1,\ldots,N_c$ and $k=1,\ldots,K$, and evaluate the achievable SINR values in \eqref{Eq:SINR1}. Figure \ref{Fig:SINRhist} displays the probability distribution of the achievable SINR values of MS$_{11}$. As seen, it is with very high chance that the achieved SINR is smaller than the target value $20$ dB, due to the presence of CSI errors. In the worst case, the actual SINR value can be even less than 5 dB. \hfill $\blacksquare$

Our goal is to design the beamforming vectors such that the SINR requirement $\gamma_{nk}$ can be {guaranteed} for all possible CSI errors. This idea {can be realized by considering} the following worst-case robust MCBF design \cite{Shen2011,Tajer11TSP}:
\begin{subequations}\label{RPM}
\begin{align}
  \min_{\substack{\wb_{nk}, k=1,\ldots,K\\ n=1,\ldots,N_c}}~&\sum_{n=1}^{N_c}{\alpha_n}\left({\sum_{k=1}^K{\left\|\wb_{nk}\right\|^2}}\right)\\
  \st~~&
  \SINR_{nk}\left(\{\wb_{m1},\ldots,\wb_{mK}\}_{m=1}^{N_c},\{\hhb_{mnk}+\eb_{mnk}\}_{m=1}^{N_c}\right)\geq \gamma_{nk}
  \notag\\
  &~\forall~\eb_{mnk}^H\Qb_{mnk}\eb_{mnk}\leq 1,~m\in\Nset,~n\in\Nset,~k\in\Kset.\label{RPM-C1}
\end{align}\end{subequations}
In comparison with the non-robust design in \eqref{PM}, the above worst-case robust MCBF design can provide guaranteed QoS for
the MSs, as illustrated in Fig. \ref{Fig:SINRhist}.
Solving the robust design problem \eqref{RPM}, however, is a challenging task. The reasons are that, firstly, each of the SINR constraints is nonconvex, and secondly, there are infinitely many such nonconvex SINR constraints due to the worst-case design criterion. 
{The restrictive approximation method presented in \cite{Tajer11TSP} is able to yield feasible approximate solutions, but is less power efficient due to the reduced problem feasible set.} In the next section, we propose a new approximation method based on relaxation, by applying the convex optimization based semidefinite relaxation (SDR) technique \cite{Luo2010_SPM}. We will further present conditions under which SDR is optimal.
Simulation results to be presented in Section \ref{Sec:Simulation} will show that the proposed SDR method is more power efficient than the method in  \cite{Tajer11TSP}.


\section{Proposed SDR based Method}\label{sec: SDR}


\subsection{Solving \eqref{RPM} by SDR and S-Lemma}\label{subsec: sdr and s-lemma}

Considering that each of the SINR constraint is nonconvex, we first apply SDR to `linearize' the robust MCBF problem \eqref{RPM}.
To illustrate this, let us express the objective function of problem \eqref{RPM} as
$\sum_{n=1}^{{N_c}}{\alpha_n}{\sum_{k=1}^K {\Tr(\wb_{nk}\wb_{nk}^H)}}$, and {express} each of the worst-case SINR
constraints in \eqref{RPM-C1} {as}
\begin{multline}\label{E0040}
\left(\hhb_{nnk}^H+\eb_{nnk}^H\right)\left(\frac{1}{\gamma_{nk}}
\wb_{nk}\wb_{nk}^H-\sum_{i\neq k}^K \wb_{ni}\wb_{ni}^H\right)
\left(\hhb_{nnk}+\eb_{nnk}\right)\notag\\~
\geq\sum_{m\neq n}^{N_c}
\left(\hhb_{mnk}^H+\eb_{mnk}^H\right)%
\left(\sum\limits_{i=1}^K\wb_{mi}\wb_{mi}^H\right)%
\left(\hhb_{mnk}+\eb_{mnk}\right)
+\sigma_{nk}^2~\forall~\eb_{mnk}^H\Qb_{mnk}\eb_{mnk}\leq
1,~m\in\Nset.
\end{multline}
The idea of SDR is to replace each rank-one matrix $\wb_{nk}\wb_{nk}^H$ by a general-rank positive semidefinite matrix
$\Wb_{nk}\succeq \zerob$, by removing the rank-one constraint on $\Wb_{nk}$ \cite{Luo2010_SPM}. After applying SDR to \eqref{RPM}, we obtain the following problem
{\small \begin{subequations}\label{intraRPM SDR}
\begin{align}
  \min_{\substack{\Wb_{nk}\succeq \zerob, k=1,\ldots,K\\ n=1,\ldots,N_c}}&\sum_{n=1}^{{N_c}}
  \alpha _n\left({\sum_{k = 1}^K{\Tr(\Wb_{nk})}}\right)\\
  \st~&
  {\left(\hhb_{nnk}^H\!+\!\eb_{nnk}^H\right)\left(\frac{1}{\gamma_{nk}}
    \Wb_{nk}\!-\!\sum_{i\neq k}^K \Wb_{ni}\right)
    \left(\hhb_{nnk}\!+\!\eb_{nnk}\right)}\geq\notag\\
    &\qquad\qquad\qquad\sum_{m\neq n}^{N_c}
    {\left(\hhb_{mnk}\!+\!\eb_{mnk}\right)^H%
     \left(\sum\limits_{i=1}^K\Wb_{mi}\right)%
     \left(\hhb_{mnk}\!+\!\eb_{mnk}\right)}
    \!+\!\sigma_{nk}^2,\notag\\
    &~\forall~\eb_{mnk}^H\Qb_{mnk}\eb_{mnk}\leq 1,~m\in\Nset,~n\in\Nset,~k\in\Kset. \label{intraRPM SDR-C2}
\end{align}
\end{subequations}} \!\!\!Note that the SDR problem \eqref{intraRPM SDR} is convex, since both the objective function and constraints are linear in $\Wb_{nk}$. However, the SDR problem \eqref{intraRPM SDR} is still computationally intractable because it involves an infinite number of constraints. Fortunately, the infinitely many constraints can be recast as {a} finite number of convex constraints.

To show this, we first observe that the left-hand side and right-hand side of the first inequality in \eqref{intraRPM SDR-C2} involve independent CSI errors. Hence, the worst-case constraint in \eqref{intraRPM SDR-C2} for $\MSnk$ can be alternatively
expressed as
\begin{align}\label{intraRPM SDR-C99}
 & \min_{\eb_{nnk}^H\Qb_{nnk}\eb_{nnk}\leq 1}\left(\hhb_{nnk}^H\!+\!\eb_{nnk}^H\right)\left(\frac{1}{\gamma_{nk}}
    \Wb_{nk}\!-\!\sum_{i\neq k}^K \Wb_{ni}\right)
    \left(\hhb_{nnk}\!+\!\eb_{nnk}\right)\geq\notag\\
    &~~~~~~~~~~~~~~~~\sum_{m\neq n}^{N_c}\bigg\{\max_{\eb_{mnk}^H\Qb_{mnk}\eb_{mnk}\leq 1}
    \left(\hhb_{mnk}\!+\!\eb_{mnk}\right)^H%
     \left(\sum\limits_{i=1}^K\Wb_{mi}\right)%
     \left(\hhb_{mnk}\!+\!\eb_{mnk}\right)\bigg\}+\sigma_{nk}^2.
\end{align}
By introducing the slack variable
\begin{align}\label{intraRPM SDR-C99-0}
   t_{mnk} = \max_{\eb_{mnk}^H\Qb_{mnk}\eb_{mnk}\leq 1}
    \left(\hhb_{mnk}\!+\!\eb_{mnk}\right)^H%
     \left(\sum\limits_{i=1}^K\Wb_{mi}\right)%
     \left(\hhb_{mnk}\!+\!\eb_{mnk}\right)
\end{align}
as the worst-case ICI power from $\BSm$ to $\MSnk$, for all $m\in\Nset\backslash\{n\}$, \eqref{intraRPM SDR-C99} can be written as
\begin{align}\label{intraRPM SDR-C99-1}
 & \min_{\eb_{nnk}^H\Qb_{nnk}\eb_{nnk}\leq 1}\left(\hhb_{nnk}^H\!+\!\eb_{nnk}^H\right)\left(\frac{1}{\gamma_{nk}}
    \Wb_{nk}\!-\!\sum_{i\neq k}^K \Wb_{ni}\right)
    \left(\hhb_{nnk}\!+\!\eb_{nnk}\right)\geq\sum_{m\neq n}^{N_c}t_{mnk}+\sigma_{nk}^2.
\end{align}
By \eqref{intraRPM SDR-C99-0} and \eqref{intraRPM SDR-C99-1}, the worst-case SINR constraint {for $\MSnk$} in \eqref{intraRPM SDR-C2} can be decoupled into the following $N_c$ worst-case constraints:
\begin{align}
&{\left(\hhb_{nnk}^H\!+\!\eb_{nnk}^H\right)
  \left(\frac{1}{\gamma_{nk}} \Wb_{nk}\!-\!\!\sum_{i\neq k}^K \Wb_{ni}\right)
  \left(\hhb_{nnk}\!+\!\eb_{nnk}\right)}
  \geq\!\sum_{m\neq n}^{N_c}t_{mnk} +\sigma_{nk}^2 ~\forall~\eb_{nnk}^H\Qb_{nnk}\eb_{nnk}\leq 1,\label{SINR_constraint2-1}\\
&{\left(\hhb_{mnk}^H+\eb_{mnk}^H\right)
\left(\sum\limits_{i=1}^K\Wb_{mi}\right)\left(\hhb_{mnk}+\eb_{mnk}\right)}\leq
t_{mnk}~ \forall~ \eb_{mnk}^H\Qb_{mnk}\eb_{mnk}\leq
1,~m\in\Nset\backslash\{n\}.\label{SINR_constraint2-2}
\end{align}
The ingredient of reformulating \eqref{SINR_constraint2-1} and \eqref{SINR_constraint2-2} into finite convex constraints is the {S}-lemma:

\begin{Lemma}[{S}-lemma\cite{BK:BoydV04}]\label{Slamma}
Let $\phi_i(\xb)\triangleq \xb^H\Ab_i\xb + \bb_i^H\xb + \xb^H\bb_i + c_i$, for $i=0,1,$ where $\Ab_i\in\mathbb{H}^{N_t}$,
$\bb_i\in\Cset^{N_t}$ and $c_i\in\Rset$. Suppose that there exits an $\bar{\xb}\in\Cset^{N_t}$ such that $\phi_1(\bar{\xb})<0$. Then the two conditions are equivalent:
\begin{enumerate}
  \item [(1)] $\phi_0(\xb)\geq 0$ for all $\xb$ satisfying $\phi_1(\xb)\leq 0$;
  \item [(2)] There exists $\lambda\geq 0$ such that
\begin{align*}
  \begin{bmatrix} \Ab_0&\bb_0\\\bb_0^H&c_0 \end{bmatrix} +
  \lambda
  \begin{bmatrix}
    \Ab_1&\bb_1\\\bb_1^H&c_1
  \end{bmatrix}\succeq \zerob.
\end{align*}
\end{enumerate}
\end{Lemma}

By applying the above {{S}}-lemma, we can equivalently recast \eqref{SINR_constraint2-1} and
\eqref{SINR_constraint2-2} as
\begin{align}
\!\!\!\Phib_{nk}&\left(\{\Wb_{ni}\}_{i=1}^K,\{t_{mnk}\}_{m},\lambda_{nnk}\right)\notag \\
&\!\!\!\! \triangleq
    \begin{bmatrix}
    \Ub_{nk}+\lambda_{nnk}\Qb_{nnk} & \Ub_{nk}\hhb_{nnk} \\
    \hhb_{nnk}^H\Ub_{nk} & \hhb_{nnk}^H\Ub_{nk}\hhb_{nnk}-\!\lambda_{nnk}\!\!-\!\!\sum\limits_{m\neq n}\!t_{mnk}\!\!-\!\sigma_{nk}^2
    \end{bmatrix}\succeq \zerob,\label{LMI1} \\
\Psib_{mnk}&\left(\{\Wb_{mi}\}_{i=1}^K, t_{mnk},\lambda_{mnk}\right) \notag\\ &\!\!\!\!\triangleq
\begin{bmatrix}
   -\left(\sum\limits_{i=1}^K \!\Wb_{mi}\right)+\lambda_{mnk}\Qb_{mnk}
&  -\left(\sum\limits_{i=1}^K \!\Wb_{mi}\right)\hhb_{mnk}\\
   -\hhb_{mnk}^H\left(\sum\limits_{i=1}^K \!\Wb_{mi}\right)
&\hspace{-5mm}
   -\hhb_{mnk}^H\left(\sum\limits_{i=1}^K \!\Wb_{mi}\right)\hhb_{mnk}+t_{mnk}\!-\!\lambda_{mnk}
\end{bmatrix}\succeq \zerob,~m\in\Nset\backslash\{n\},\label{LMI2}
\end{align}
where
$\Ub_{nk}\triangleq \frac{1}{\gamma_{nk}}\Wb_{nk}-\sum_{i\neq k}^K \Wb_{ni}$, and $\lambda_{mnk}\geq 0$ are slack variables.
In summary, one can reformulate \eqref{intraRPM SDR} as the following problem
\begin{subequations}\label{RobustMCBF}
\begin{align}
\min_{\substack{\{\Wb_{nk}\},\\\{\lambda_{mnk}\},\{t_{mnk}\}}}~
  &\sum_{n=1}^{N_c}\alpha_n \left(\sum_{k=1}^K\Tr(\Wb_{nk})\right)\\
  \st~
  &\Phib_{nk} \left(\{\Wb_{ni}\}_{i=1}^K,\{t_{mnk}\}_{m},\lambda_{nnk}\right) \succeq \zerob~ \forall~n\in\Nset,~k\in \Kset \label{RobustMCBF-b1}\\
  &\Psib_{mnk}\left(\{\Wb_{mi}\}_{i=1}^K,  t_{mnk}, \lambda_{mnk}\right) \succeq \zerob
  ~\forall~m\in\Nset\backslash\{n\},~n\in\Nset,~k\in \Kset, \label{RobustMCBF-c1}\\
  &\Wb_{nk}\succeq \zerob,~\lambda_{mnk}\geq 0~\forall~m,n\in\Nset,~k\in \Kset.\label{RobustMCBF-e1}
\end{align}
\end{subequations}
Problem \eqref{RobustMCBF} is a convex semidefinite program (SDP) which can be efficiently solved {by off-the-shelf convex solvers \cite{SeDuMi}}.


\subsection{Optimality Conditions}

An important aspect of SDR is whether the relaxed problem can yield a rank-one solution, i.e., whether the optimal solution $\{\Wb_{nk}^{\star}\}$ satisfies $\Wb_{nk}^{\star}=\wb_{nk}^{\star}(\wb_{nk}^{\star})^H$ for some $\wb_{nk}^\star\in \mathbb{C}^{N_t}$, for all $n$, $k$. If this is true, then $\{\wb_{nk}^{\star}\}$ is an optimal solution of the original robust MCBF problem \eqref{RPM}. It, therefore, is important to investigate the conditions under which the SDR
problem \eqref{RobustMCBF} can yield a rank-one solution. Some provable conditions are given in the following
proposition:

\begin{Prop} \label{prop: rank-one}
Suppose that the SDR problem \eqref{RobustMCBF} is feasible. Consider the following three conditions:
\begin{enumerate}
\item [{C1)}] $K=1$, i.e., there is only one MS in each cell;

\item [{C2)}] $\Qb_{nnk}=\infty\Ib_{N_t}$ for all $n$, $k$, i.e., $\eb_{nnk}=\zerob$ for all $n$, $k$, and thus perfect intra-cell CSI $\{\hb_{nnk}\}$; 

\item [{C3)}] For the spherical error model, i.e.,
$\|\eb_{mnk}\|^2 \leq\varepsilon_{mnk}^2~\text{for all}~m,n,k,$ the CSI error bounds $\{\varepsilon_{mnk}\}$ satisfy
\begin{align}\label{conditions 1}
   \varepsilon_{mnk} \leq \bar\varepsilon_{mnk}~\text{and}~
   \varepsilon_{nnk} < \sqrt{\frac{\sigma_{nk}^2\alpha_n\gamma_{nk}}{f^\star}}
\end{align}
for all $m,n,k$, where $\{\bar\varepsilon_{mnk}^2\}$ are some CSI error bounds under which problem \eqref{RobustMCBF} is feasible, with $f^\star>0$ denoting the associated optimal objective value.
\end{enumerate}
If any one of the above three conditions is satisfied, then the SDR problem \eqref{RobustMCBF} must yield a rank-one solution,
that is, the optimal solution, denoted by $\{\Wb_{nk}^\star\}$, must satisfy
  \[\Wb_{nk}^\star=\wb_{nk}^\star(\wb_{nk}^\star)^H~\text{for all}~n,k.\]
\end{Prop}


The proof of Proposition \ref{prop: rank-one} is presented in Appendix \ref{proof of rank one}. Case C1) of Proposition \ref{prop: rank-one} shows that the global optimum of the robust MCBF problem \eqref{RPM} can be attained by solving the SDR problem \eqref{RobustMCBF} whenever $K=1$.
We should mention that, when $K=1$, problem \eqref{RPM} can be handled alternatively by a convex conic reformulation approach presented in \cite{Tajer11TSP}.
For the general case of $K>1$, {C2) shows that if the BSs have channel uncertainty only for inter-cell CSI, i.e., $\{\hb_{mnk}\}$ where $m\neq n$, and perfectly know
the intra-cell CSI, i.e., $\{\hb_{nnk}\}$}, then rank-one solutions are guaranteed. If errors occur in both intra-cell and inter-cell CSI, C3) states that rank-one solutions can also be obtained by solving the SDR problem \eqref{RobustMCBF}, provided that the CSI errors are sufficiently small.

For a general setup, it is not known theoretically whether the SDR problem \eqref{RobustMCBF} has a rank-one solution. If the obtained solution is not of rank one, then additional solution approximation procedure, such as the Gaussian randomization method
\cite{BK:LuoChang}, can be used for obtaining a rank-one approximate solution to problem \eqref{RPM}. Quite surprisingly, we found in our simulation tests (see Section \ref{subsec:Simulation setting} for the setting), that problem \eqref{RobustMCBF} always yields rank-one optimal
$\{\Wb_{nk}^{\star}\}$. 
Hence for these problem instances, we can simply perform rank-one decomposition of $\Wb_{nk}^{\star}=\wb_{nk}^{\star}(\wb_{nk}^{\star})^H$. Investigating the reasons behind would be an interesting future research; see \cite{Song2011,Chang2011Asilomar} for recent endeavors.

\section{Distributed Robust MCBF Algorithm using ADMM} \label{Sec:RobustDMCBF}

Solving the SDR problem \eqref{RobustMCBF} calls for a control center which computes the beamforming solutions
in a centralized manner using all the CSI of MSs. As discussed in the introduction, it is desirable to
obtain the beamforming solutions in a decentralized fashion using local CSI at each BS, i.e., $\BSn$ uses $\{\hhb_{nmi}\}_{m,i}$ only, for $n=1,\ldots,N_c$.
A simple {approach would} be applying the dual decomposition method \cite{Boyddecomposition}, similar to the works in \cite{Tolli11TWC,Tajer11TSP}. However, as will be explained later, the dual decomposition method is not suitable for problem \eqref{RobustMCBF} due to the fact that the decomposed problems lack strict convexity and may be unbounded below.
%
%
To fix this, we propose the use of \emph{alternating direction method of multipliers} (ADMM) \cite{BertsekasADMM,BoydADMM} for distributed optimization of the SDR problem \eqref{RobustMCBF}.
%
In the first subsection, we briefly review ADMM.
In the second subsection, we show how an effective distributed robust MCBF algorithm can be developed following the principle of ADMM.

\vspace{-0.2cm}
\subsection{Review of ADMM}\label{ADMM}

To illustrate the idea of ADMM, let us consider the following convex optimization problem\cite{BertsekasADMM,BoydADMM}
\vspace{-0.5cm}
\begin{subequations}\label{Prob:ADMM}
\begin{align}
\min_{\xb\in \Rset^n,\zb\in \Rset^m}~&F(\xb)+G(\zb)\\
\st~~&\xb\in \mathcal{S}_1,~\zb\in \mathcal{S}_2, \\
&\zb=\Ab\xb,
\end{align}
\end{subequations}
where $F:\Rset^n\mapsto\Rset$ and $G:\Rset^m\mapsto\Rset$ are convex functions, $\Ab$ is an $m\times n$ matrix, and $\mathcal{S}_1\subset\Rset^n$ and $\mathcal{S}_2\subset\Rset^m$ are nonempty polyhedral sets. Assume that problem \eqref{Prob:ADMM} is solvable and strong duality holds.

ADMM considers the following penalty augmented problem
\begin{subequations}\label{Prob:ADMM2}
\begin{align}
\min_{\xb\in \Rset^n,\zb\in \Rset^m}~&F(\xb)+G(\zb)+\frac{c}{2}\|\Ab\xb-\zb\|^2\\
\st~~&\xb\in \mathcal{S}_1,~\zb\in \mathcal{S}_2, \\
&\zb=\Ab\xb, \label{Prob:ADMM2 C3}
\end{align}
\end{subequations}
where $c>0$ is the penalty parameter. It is easy to see that \eqref{Prob:ADMM2} is essentially equivalent to \eqref{Prob:ADMM} owing to \eqref{Prob:ADMM2 C3}.
The penalty term $\frac{c}{2}\|\Ab\xb-\zb\|^2$ brings strict convexity; as seen, problem \eqref{Prob:ADMM2} is strictly convex with respect to either
$\xb$ or $\zb$.

The second ingredient of ADMM is dual decomposition \cite{BoydADMM} where the dual problem of \eqref{Prob:ADMM2} is concerned:
\begin{align}\label{MoM15}
    \max_{\xib\in \Rset^m}
    \left\{
    \begin{array}{rl}
        \min\limits_{\xb,\zb} & F(\xb)+G(\zb)+\frac{c}{2}\|\Ab\xb-\zb\|^2 +\xib^T(\Ab\xb-\zb)\\
        \st  &\xb\in \mathcal{S}_1,~\zb\in \mathcal{S}_2
    \end{array}
    \right\}
\end{align} in which $\xib\in \Rset^m$ is the dual variable associated with the constraint \eqref{Prob:ADMM2 C3}. 
Given a dual variable $\xib$, the inner problem is a convex problem and can be efficiently solved. The outer variable $\xib$ can be updated by the subgradient method \cite{BK:BoydV04}. In a standard dual optimization procedure, one usually updates the outer variable $\xib$ when the associated inner problem has been solved with the global optimum. For example, one can use the nonlinear Gauss-Seidel method \cite{BertsekasADMM} to optimally solve the inner problem. Specifically, in the nonlinear Gauss-Seidel method, one iteratively solves the following two subproblems
\begin{subequations}\label{Gauss seidal}
\begin{align}
\zb(q+1) &=\arg\min\limits_{\zb\in \mathcal{S}_2} \left\{G(\zb)-\xib(q)^T\zb+\frac{c}{2}\|\Ab\xb(q)-\zb\|^2\right\}, \\
\xb(q+1) &=\arg\min\limits_{\xb\in \mathcal{S}_1} \left\{F(\xb)+\xib(q)^T\Ab\xb+\frac{c}{2}\|\Ab\xb-\zb(q+1)\|^2\right\}
\end{align}
\end{subequations}
until convergence, where $q$ is the iteration number. Instead, ADMM, as its name suggests, alternatively performs one iteration of the Gauss-Seidel step \eqref{Gauss seidal} and one step of the outer subgradient update for speeding up its convergence. The steps of ADMM are summarized in Algorithm 1.

\begin{algorithm}[h]
\caption{\textbf{ADMM}}
\begin{algorithmic}[1]\label{table: ADMM}
\STATE Set $q=0$, choose $c>0$, %
\STATE Initializes $\xib(q)$ and $\xb(q)$;

\REPEAT

\STATE $\zb(q+1) =\arg\min\limits_{\zb\in \mathcal{S}_2} \left\{G(\zb)-\xib^T(q)\zb+\frac{c}{2}\|\Ab\xb(q)-\zb\|^2\right\}$

\STATE $\xb(q+1) =\arg\min\limits_{\xb\in \mathcal{S}_1} \left\{F(\xb)+\xib^T(q)\Ab\xb+\frac{c}{2}\|\Ab\xb-\zb(q+1)\|^2\right\}$

\STATE $\xib(q+1)=~\xib(q)+c\left(\Ab\xb(q+1)-\zb(q+1)\right)$

\STATE $q:=q+1$;%
\UNTIL the predefined stopping criterion is satisfied.%
\end{algorithmic}
\end{algorithm}\vspace{-0cm}

ADMM can actually converge to the global optimum of problem \eqref{Prob:ADMM} under relatively loose conditions; see the following lemma from \cite[Proposition 4.2]{BertsekasADMM}:

\begin{Lemma}{\rm \cite[Proposition 4.2]{BertsekasADMM}}\label{ADMMConvergence}
Assume that $\mathcal{S}_1$ is bounded or that $\Ab^T\Ab$ is invertible. A sequence $\{\xb(q),\zb(q),\xib(q)\}$ generated by Algorithm \ref{table: ADMM} is bounded, and every limit point of $\{\xb(q),\zb(q)\}$ is an optimal solution of the original problem \eqref{Prob:ADMM}.
\end{Lemma}

\subsection{Applying ADMM to Problem \eqref{RobustMCBF}}\label{subsec: distributed algorithm}

Applying ADMM to our SDR problem \eqref{RobustMCBF} for distributed optimization is not as straightforward because the constraints in \eqref{RobustMCBF} are intricately coupled. Our intension in this subsection is to reformulate problem \eqref{RobustMCBF} such that the corresponding ADMM steps in Algorithm \ref{table: ADMM} are decomposable and thus problem \eqref{RobustMCBF} can be solved in a distributed fashion.

To this end, we first introduce the following two auxiliary variables
\begin{align}
  p_n=\sum\limits_{k=1}^K\Tr(\Wb_{nk}),~T_{nk}=\sum\limits_{m\neq n}^{N_c}t_{mnk}
\end{align} for all $n\in \Nset$ and $k\in \Kset$, where $p_n$ represents the
transmission power of $\BSn$,
and $T_{nk}$ stands for the total worst-case ICI power from the neighboring BSs
to $\MSnk$. Then problem \eqref{RobustMCBF} can be rewritten as
\begin{subequations}\label{intraSDP_0010}
\begin{align}
\min_{\substack{\{\Wb_{nk}\succeq\zerob\},\\\{\lambda_{mnk}\geq 0\},\{t_{mnk}\}\\ \{p_n\},\{T_{nk}\}}}~
  &\sum_{n=1}^{N_c}\alpha_n p_n \\
  \st~
  &\Phib_{nk} \left(\{\Wb_{ni}\}_{i=1}^K,T_{nk},\lambda_{nnk}\right) \succeq \zerob~\forall~n\in\Nset,~k\in \Kset,\label{intraSDP_0010 C1}\\
  &\Psib_{mnk}\left(\{\Wb_{mi}\}_{i=1}^K,  t_{mnk}, \lambda_{mnk}\right) \succeq \zerob,
  ~\forall~m\in\Nset\backslash\{n\},~n\in\Nset,~k\in \Kset, \label{intraSDP_0010 C2} \\
  &\sum\limits_{k=1}^K\Tr(\Wb_{nk})=p_n~\forall~ n\in\Nset,\label{intraSDP_0010 C4}\\
  &\sum\limits_{m\neq n}^{N_c}t_{mnk}=T_{nk}~\forall~ n\in\Nset,~k\in \Kset. \label{intraSDP_0010 C5}
\end{align}
\end{subequations}
It is interesting to observe from \eqref{intraSDP_0010 C1} that each $\MSnk$ concerns only the total worst-case ICI power $T_{nk}$
instead of the individual worst-case ICI powers $\{t_{mnk}\}$. As will be clear later, introducing such slack variables $\{T_{nk}\}$ will reduce the backhaul signaling overhead for distributed optimization.

Note that, without loss of generality, we can interchange the subindices $m$ and $n$ in \eqref{intraSDP_0010 C2}.
Hence, the constraints in \eqref{intraSDP_0010 C1} to \eqref{intraSDP_0010 C4} can be decomposed into $N_c$ independent convex sets:
\begin{multline}\label{Cn}
\cC_n=\bigg\{\!\!\bigg(\{\Wb_{nk}\}_k,\{\lambda_{nmk}\}_{m,k},\{T_{nk}\}_k,\{t_{nmk}\}_{m,k},p_n\bigg)\bigg|\bigg.\!\!\!\! \bigg.\\
 \begin{array}{ll}
        &\Phib_{nk} \left(\{\Wb_{ni}\}_{i=1}^K,T_{nk},\lambda_{nnk}\right) \succeq \zerob~\forall~ k\in \Kset,\\
        &\Psib_{nmk}\left(\{\Wb_{ni}\}_{i=1}^K,  t_{nmk}, \lambda_{nmk}\right)
        \succeq \zerob~\forall~m\in\Nset\setminus\{n\},~k\in \Kset,\\
      &\lambda_{nmk}\geq 0~\forall~m\in\Nset,~k\in \Kset,\\
     &\bigg.\sum\limits_{k=1}^K\Tr(\Wb_{nk})=p_n,~\Wb_{nk}\succeq \zerob~\forall~k\in \Kset\bigg\}
\end{array}
\end{multline}
\!\! for $n=1,\ldots,N_c$. 
Further define the following variables
\begin{subequations}\label{notation010}
\begin{align}\label{t tn}
\tb&=    \begin{bmatrix}t_{121},\ldots,t_{12K},~\ldots,~t_{N_c(N_c-1)1},\ldots,t_{N_c(N_c-1)K}\end{bmatrix}^T \in \Rset^{Nc(Nc-1)K}_+,\\
\tb_n&=\begin{bmatrix}T_{n1},\ldots,T_{nK}, &t_{n11},\ldots,t_{n1K},t_{n21},\ldots,t_{nN_cK}\end{bmatrix}^T\in \Rset^{NcK}_+,~n=1,\ldots,N_c,
\end{align}
\end{subequations}
where $\tb$ collects all the ICI variables, and $\tb_n$ collects variables $\{T_{nk}\}_{k=1}^K$ and $\{t_{nmk}\}_{m,k}$ (where $m\neq n$) that are
relevant only to $\BSn$. It is not difficult to check that there {exists a linear mapping matrix $\Eb_n\in\Rset^{N_cK\times N_c(N_c-1)K}$, which {contains only} elements either equal to one or zero, such that
\begin{align}\label{En}
   \tb_n=\Eb_n\tb,
\end{align}
for all $n=1,\ldots,N_c$.} %
%
By \eqref{Cn}, \eqref{t tn} and \eqref{En}, we can rewrite problem \eqref{intraSDP_0010} in a compact form as
\begin{subequations}\label{intraSDP_0020}
\begin{align}
\min_{\{\Wb_{nk}\},\{\lambda_{nmk}\},\{\tb_n\},\{p_n\},\tb} ~& \sum_{n=1}^{N_c} \alpha_n p_n\\
\st~~ &(\{\Wb_{nk}\}_k,\{\lambda_{nmk}\}_{m,k},\tb_n,p_n)\in\cC_n,~ n=1,\ldots,N_c,\\
&\tb_n=\Eb_n\tb,~ n=1,\ldots,N_c.\label{intraSDP_0020 C3}
\end{align}
\end{subequations}

Before applying ADMM, let us first see why the conventional dual decomposition method \cite{Boyddecomposition} is not suitable {for} problem \eqref{intraSDP_0020}. Consider the dual problem of \eqref{intraSDP_0020}
\begin{align}\label{dual of intraSDP_0020}
\max_{\substack{\nub_n\in \Rset^{N_cK}, \nub_n^H\Eb_n=\zerob,\\ n=1,\ldots,N_c }}
\left\{
\min_{\substack{(\{\Wb_{nk}\}_k,\{\lambda_{nmk}\}_{m,k},\tb_n,p_n)\in\cC_n,\\n=1,\ldots,N_c}} ~
\sum_{n=1}^{N_c} \alpha_n p_n - \sum_{n=1}^{N_c} \nub_n^T\tb_n
\right\},
\end{align}where $\nub_n\in \mathbb{R}^{N_cK}$, $n=1,\ldots,N_c$, are the dual variables associated with constraints \eqref{intraSDP_0020 C3}.
While the inner minimization problem of \eqref{dual of intraSDP_0020} is obviously decomposable, given $\nub_1,\ldots,\nub_{N_c}$, it is possible for one to obtain an inner solution of $\tb_n$ such that $-\nub_n^T\tb_n \rightarrow -\infty$, i.e.,
the inner minimization problem is unbounded below, due to the unboundedness of the feasible sets $\cC_n$ {(see \eqref{Cn}, \eqref{LMI1} and \eqref{LMI2})}.
In fact, our numerical experience shows
that this undesired situation happens very often, especially when $N_c>2$.

To overcome this issue, we apply the augmented Lagrangian method to \eqref{intraSDP_0020} according to the principle of ADMM; this leads to
the following problem:
\begin{subequations}\label{intraSDP_0060}
\begin{align}
\min_{\substack{\{\Wb_{nk}\},\{\lambda_{nmk}\},\{\tb_n\},\\ \{p_n\},\{\rho_n\},\tb}} ~& \sum_{n=1}^{N_c} \alpha_n p_n+\frac{c}{2}\sum_{n=1}^{N_c}\left\|\Eb_n\tb-\tb_n\right\|^2
 +\frac{c}{2}\sum_{n=1}^{N_c}\left(\rho_n-p_n\right)^2 \label{intraSDP_0060 OBF}\\
\st~~ &(\{\Wb_{nk}\}_k,\{\lambda_{nmk}\}_{m,k},\tb_n,p_n)\in\cC_n, n=1,\ldots,N_c,\\
&\tb_n=\Eb_n\tb, n=1,\ldots,N_c, \label{intraSDP_0060 C2}\\
&p_n=\rho_n, n=1,\ldots,N_c,\label{intraSDP_0060 C3}
\end{align}
\end{subequations}
where $\rho_n\geq 0$, $n=1,\ldots,N_c$, are slack variables, which are introduced {in order to impose the penalty term $\frac{c}{2}\sum_{n=1}^{N_c}\left(\rho_n-p_n\right)^2$}. Problem \eqref{intraSDP_0060} is equivalent to problem \eqref{intraSDP_0020}, but the added penalty terms {can} resolve
the numerically unbound below {issue}; see \cite{BertsekasADMM}.

Now we are ready to apply ADMM. Consider the following {correspondences} between \eqref{intraSDP_0060} and \eqref{Prob:ADMM2}:
\begin{align}
  & \label{A} \xb\triangleq [\tb^T,\rho_1,\ldots,\rho_{N_c}]^T,~
  \zb\triangleq [\tb_1^T,\ldots,\tb_{N_c}^T,p_1,\ldots,p_{N_c}]^T,~
  \Ab\triangleq \begin{bmatrix}
   \Eb & \zerob \\
   \zerob & \Ib_{N_c}
  \end{bmatrix}, \\
  &F(\xb)\triangleq 0,~G(\zb)\triangleq \sum_{n= 1}^{N_c}\alpha_n p_n,
  \notag \\
  &\mathcal{S}_1\triangleq \Cset^{N_c(N_c-1)K+N_c}, \notag \\
  &\mathcal{S}_2\triangleq \left\{[\tb_1^T,\ldots,\tb_{N_c}^T,p_1,\ldots,p_{N_c}]^T|(\{\Wb_{nk}\}_k,\{\lambda_{nmk}\}_{m,k},
  \tb_n,p_n)\in\cC_n,n= 1,\ldots,N_c\right\}, \notag
  \\
  & \xib \triangleq  [\nub_1^T,\ldots,\nub_{N_c}^T,\mu_1,\ldots,\mu_{N_c}]^T,\notag
\end{align}
where $\Eb  \triangleq \begin{bmatrix}\Eb_1^T &\cdots&\Eb_{N_c}^T \end{bmatrix}^T$, and $\nub_n\in \mathbb{R}^{N_cK}$, $\mu_n\in \Rset$, $n=1,\ldots,N_c$, are the dual variables associated with constraints \eqref{intraSDP_0060 C2} and \eqref{intraSDP_0060 C3}, respectively.
According to Algorithm \ref{table: ADMM}, the corresponding ADMM step 4 for problem \eqref{intraSDP_0060} is to solve the following problem
\begin{align}\label{update tb pn0}
\min_{\substack{\{\Wb_{nk}\}_k,\{\lambda_{nmk}\}_{m,k},\tb_n,p_n\\n=1,\ldots,N_c}} ~ &\sum_{n=1}^{N_c}\left(\alpha_n p_n+\frac{c}{2}\left\|\Eb_n\tb(q)-\tb_n\right\|^2
 +\frac{c}{2}\left(\rho_n(q)-p_n\right)^2-\nub_n^T(q)\tb_n - \mu_n(q)p_n\right) \notag \\
\st&~ (\{\Wb_{nk}\}_k,\{\lambda_{nmk}\}_{m,k},\tb_n,p_n)\in\cC_n,~n=1,\ldots,N_c.
\end{align}
As one can see, problem \eqref{update tb pn0} can be decomposed as the following $N_c$ problems:
\begin{align}\label{update tb pn}
&\{\tb_n(q+1),p_n(q+1)\}=\notag \\
&~~~~\arg~\min_{\{\Wb_{nk}\}_k,\{\lambda_{nmk}\}_{m,k},\tb_n,p_n} ~ \alpha_n p_n+\frac{c}{2}\left\|\Eb_n\tb(q)-\tb_n\right\|^2
 +\frac{c}{2}\left(\rho_n(q)-p_n\right)^2-\nub_n^T(q)\tb_n - \mu_n(q)p_n \notag \\
&~~~~~~~~~~~~~~~~~~~~~~~~~~~~~~~~~\st~~ (\{\Wb_{nk}\}_k,\{\lambda_{nmk}\}_{m,k},\tb_n,p_n)\in\cC_n,
\end{align}
for $n=1,\ldots,N_c.$ Since problem \eqref{update tb pn} is convex, it can be efficiently solved.

Secondly, the corresponding ADMM step 5 is given by solving the following two problems:
\begin{align}\label{step 4 1}
  \tb(q+1)&=\arg~\min_{\tb\in \Rset^{N_c(N_c-1)K}_+} \frac{c}{2}\sum_{n=1}^{N_c}\|\Eb_n\tb-\tb_n(q+1)\|^2+\sum_{n=1}^{N_c}\nub_{n}^T(q)\Eb_n\tb, \\
  \{\rho_n(q+1)\}_{n=1}^{N_c}&=
  \arg~\min_{\substack{\rho_n\geq 0,\\n=1,\ldots,N_c}}~ \frac{c}{2}\sum_{n=1}^{N_c}(\rho_n-p_n(q+1))^2+
  \sum_{n=1}^{N_c}\mu_n(q)\rho_n.\label{step 4 2}
\end{align}
Because both \eqref{step 4 1} and \eqref{step 4 2} are convex quadratic problems, they have {closed-form solutions given by}
\begin{subequations}\label{update t p}
\begin{align}
\tb(q+1)&
=\Eb^\dag\left(\tilde\tb(q+1)-\frac{1}{c}\tilde\nub(q)\right),
\label{update_t}\\
\rho_n(q+1)&=p_n(q+1)-\frac{1}{c}\mu_n(q),~n=1,\ldots,N_c,\label{update_p}
\end{align}
\end{subequations}
where $\tilde\tb(q+1)=[\tb_1^T(q+1),\ldots,\tb_{N_c}^T(q+1)]^T$ and $\tilde\nub(q)=[\nub_1^T(q),\ldots,\nub_{N_c}^T(q)]^T$.

Finally, the corresponding ADMM step 6 is given by the following dual variable update
\begin{subequations}\label{intraSDP_0140}
\begin{align}\label{update nu}
\nub_n(q+1)&=\nub_n(q) + c\left(\Eb_n\tb(q+1)-\tb_n(q+1)\right),\\
\mu_n(q+1) &=\mu_n(q)  + c\left(\rho_n(q+1)-p_n(q+1)   \right),\label{update mu}
\end{align}
\end{subequations}
for $n=1,\ldots,N_c$.

It is important to note that the ADMM steps \eqref{update tb pn}, \eqref{update t p} and \eqref{intraSDP_0140} can be implemented in a distributed fashion.
Essentially, given the knowledge of local CSI $\{\hhb_{nmk}\}_{m,k}$, the optimization problem \eqref{update tb pn} can be independently solved by $\BSn$, for all $n=1,\ldots,N_c$.
After that, each $\BSn$  broadcasts its new $\tb_n$ to the other BSs. With the knowledge of $\{\tb_n\}$, each BS can compute the public variable $\tb$ by step \eqref{update_t}. Moreover, $\rho_n(q+1)$, $\nub_n(q+1)$ and $\mu_n(q+1)$
in \eqref{update_p}, \eqref{update nu} and \eqref{update mu} all can be independently updated by each $\BSn$, for $n=1,\ldots,N_c$.
Summarizing the above steps, we thus obtain the distributed robust MCBF algorithm in Algorithm \ref{RobustDMCBF}.

\begin{algorithm}[h]
\caption{\textbf{Proposed Distributed Robust MCBF Algorithm}:}
\begin{algorithmic}[1]\label{RobustDMCBF}
\STATE {\bf Input} a set of initial variables
$\{\nub_n(0),\mu_n(0),\tb(0),\rho_n(0)\}_{n=1}^{N_c}$ that are known to all BSs;
choose a penalty parameter $c>0$.
\STATE Set $q=0$.
\REPEAT

\STATE Each $\BSn$ solves the local beamforming design problem \eqref{update tb pn}
to obtain the local ICI variables $\tb_n(q\!+\!1)$ and the local power $p_n(q\!+\!1)$.
\STATE Each $\BSn$ broadcasts its local ICI variable $\tb_n$ to the other BSs, e.g., via the backhaul network.

\STATE Each $\BSn$ updates the public ICI $\tb$ and $\rho_n$ by \eqref{update_t} and \eqref{update_p}, respectively.

\STATE Each $\BSn$ updates the dual variable $\nub_n$ an $\mu_n$ by \eqref{update nu}
and \eqref{update mu}, respectively.
\STATE Set $q:=q+1$;%
\UNTIL {the predefined stopping criterion is met.}
\end{algorithmic}
\end{algorithm}

Algorithm 2 is guaranteed to converge to the global optimum of the SDR problem \eqref{RobustMCBF}. Specifically, one can verify that the matrix $\Ab$ in \eqref{A}
satisfies
\begin{align}\label{AA invertible}
  \Ab^T\Ab=\begin{bmatrix}\sum_{n=1}^{N_c}\Eb_n^T\Eb_n& \zerob \\ \zerob &\Ib_{N_c} \end{bmatrix} \succ \zerob.
\end{align} 
By Lemma \ref{ADMMConvergence} and by the fact of
\eqref{AA invertible}, we obtain the following result on the convergence of Algorithm 2:

\begin{Prop}\label{prop: convergence} Consider the proposed distributed robust MCBF algorithm
in Algorithm \ref{RobustDMCBF}. The iterates $\tb(q)$, $\{p_n(q),\tb_n(q),\rho_n(q)\}_{n=1}^{N_c}$
and $\{\nub_n(q),\mu_n(q)\}_{n=1}^{N_c}$ will respectively converge to the optimal primal
and dual solutions of problem \eqref{intraSDP_0060} as $q\rightarrow \infty$. When the
algorithm converges, the optimal $\{\Wb_{n1},\ldots,\Wb_{nK}\}_{n=1}^{N_c}$ obtained
in Step 4 is a global optimal solution of the SDR problem \eqref{RobustMCBF}.
\end{Prop}

Three remarks regarding the proposed distributed robust MCBF algorithm are in order.


\vspace{-0.2cm}
\begin{Remark}\rm
In Algorithm 2, each $\BSn$ needs to exchange its local ICI power vector $\tb_n(q+1)$ with the other BSs through
the backhaul network. Since $\tb_n(q+1)$ contains $K$ total incoming ICI powers $\{T_{nk}\}_{k=1}^{N_c}$ and $(N_c-1)K$ outgoing ICI powers $\{t_{nmk}\}_{m\neq n,k}$, the total backhaul signaling is $N_c^2K$ real variables for each iteration. {By contrast, for the distributed algorithms in \cite{Tolli11TWC,Tajer11TSP}, each BS has to exchange $(N_c-1)K$ real variables for incoming ICI powers and $(N_c-1)K$ real variables for outgoing ICI powers, and thus a total number of $2N_c(N_c-1)K$ real variables need to be exchanged} for each iteration. For $N_c>2$, the proposed algorithm is clearly more
backhaul bandwidth efficient.
For example, if $N_c=6$, the required backhaul signaling of the proposed algorithm is {about} $60\%$ of that of the algorithms in \cite{Tolli11TWC,Tajer11TSP}.
\end{Remark}

\vspace{-0.4cm}
\begin{Remark}\rm
{Interestingly, Algorithm 2 can be interpreted as an adaptive ICI regularization scheme} where the cooperative BSs gradually
control their own beamforming solutions until a consensus on the induced ICI powers among BSs is reached.
To further explain it, Step 6 of the algorithm can be regarded as a step that computes the tentatively
consentient ICI power vector $\tb(q+1)$ based on the locally optimized ICI power vectors $\{\tb_n(q)\}_{n=1}^{N_c}$.
In Step 7, the BSs then update the dual variables according to the difference between the consentient ICI powers and
local ICI powers. Once the algorithm converges (which implies $\Eb_n\tb(q+1)=\tb_n(q+1)$ for all $n$), all the BSs
achieve a global consensus on the ICI powers, and hence the beamforming solutions obtained in Step 4 are globally optimal
(by Proposition \ref{prop: convergence}).
\end{Remark}

\vspace{-0.4cm}
\begin{Remark}\rm
Since ADMM operates in the dual domain, the obtained $\{\Wb_{nk}\}$ and $\{\lambda_{nmk}\}$ in Step 4 may not be feasible to the primal SDR problem \eqref{RobustMCBF}. To fix this, each BS may perform one more primal optimization of
\begin{align}\label{Prob:DistLast}
\min\limits_{{\{\Wb_{nk}\succeq \zerob\}_k},{\{\lambda_{nmk}\geq 0\}_{m,k}}}\sum\limits_{k=1}^K~&\Tr(\Wb_{nk})\\
\st~&
\Phib_{nk} \left(\{\Wb_{ni}\}_{i=1}^K,\{t_{mnk}(q+1)\}_{m\neq n},\lambda_{nnk}\right) \succeq \zerob~\forall~ k\in \Kset,\notag\\
&\Psib_{nmk}\left(\{\Wb_{ni}\}_{i=1}^K, t_{nmk}(q+1),\lambda_{nmk}\right) \succeq \zerob~\forall~ m\in \Nset\setminus \{n\},~k\in \Kset,\notag
\end{align}
using the tentatively consented ICI power vector $\tb(q+1)$. The obtained $\{\Wb_{nk}\}$ and $\{\lambda_{nmk}\}$ then must be feasible to the SDR problem \eqref{RobustMCBF}, provided that problem \eqref{Prob:DistLast} is feasible for all BSs. If at least one of the BSs declares infeasibility of \eqref{Prob:DistLast}, then more iterations are needed for Algorithm 2 since the algorithm may stop too early to reach a reasonable consensus on the global ICI $\tb(q+1)$.
\end{Remark}

\section{Extension to Fully Coordinated BSs}\label{sec:edge user}

In this section, we extend the robust MCBF design to the scenario where some of the MSs are served simultaneously by multiple BSs.
The scenario may occur, for example, when some of the MSs are near the cell boundary and thus desire to receive the information signal sent from multiple BSs {for the guaranteed QoS.}
To simultaneously serve these MSs, the BSs have to be fully coordinated, with shared data streams and CSI of these cell-edge MSs \cite{Gesbert10JSAC}.
Assume that there are $L$ such cell-edges MSs, in addition to the $N_cK$ intra-cell MSs that are served solely by their respective BSs.
The transmit signal of $\BSn$ is given by
\begin{align}\label{eq. transmit signal edgeuser}
  \tilde\xb_n(t)=\xb_n(t)+\sum_{\ell=1}^L{\fb_{n\ell}d_{\ell}(t)},
\end{align} where $\xb_n(t)$ is defined in \eqref{eq. transmit signal}
which is intended for the $K$ intra-cell MSs,
$d_{\ell}(t)$ is the data stream for the $\ell$th cell-edge MS, and $\fb_{n\ell}\in \Cset^{N_t}$ is the beamforming vector of $\BSn$ for sending $d_{\ell}(t)$. The received signals of intra-cell $\MSnk$ and cell-edge MS $\ell$ are respectively given by
\begin{align}\label{eq. received signal edgeuser}
y_{nk}^{\rm (Intra )}(t)
=&\sum_{m=1}^{N_c}\hb_{mnk}^H\xb_m(t-\tau_{nk}^{(m)}) + \sum_{m=1}^{N_c}\sum_{j=1}^{L}\hb_{mnk}^H\fb_{nj}d_{j}(t-\tau_{nk}^{(m)})+z_{nk}(t),
\end{align}
\begin{align}
y_{\ell}^{\rm (Edge)}(t)
=&\sum_{m=1}^{N_c}\gb_{m\ell}^H \xb_m(t-\tau_{\ell}^{(m)}) + \sum_{m=1}^{N_c}\sum_{j=1}^{L}\gb_{m\ell}^H\fb_{nj}d_{j}(t-\tau_{\ell}^{(m)})+z_{\ell}(t),\label{eq. received signal edgeuser2}
\end{align}
for $n\in \Nset$, $k\in \Kset$ and $\ell\in\Lset\triangleq \{1,\ldots,L\}$,
where $\gb_{m\ell}\in \Cset^{N_t}$ is the channel vector from $\BSm$ to cell-edge MS $\ell$, and $z_{\ell}(t)$ is the background noise at cell-edge MS $\ell$, which is assumed to be zero mean and with {variance} $\sigma_\ell^2>0$.
Note from \eqref{eq. received signal edgeuser} and \eqref{eq. received signal edgeuser2} that
{we have taken into account} the {inevitable time
delays} $\tau_{nk}^{(m)}, \tau_{\ell}^{(m)}>0$ between the BSs and MSs \cite{Zhang08TWC,Bjornson2010}. 
{Assume that $\tau_{\ell}^{(m)}\neq \tau_{\ell}^{(n)}$ for all $m\neq n$, and that each $d_{\ell}(t)$ is temporally
uncorrelated with zero mean and unit variance.} The receiver SINRs corresponding to
\eqref{eq. received signal edgeuser} and \eqref{eq. received signal edgeuser2} are given by \cite{Bjornson2010}
\begin{align}\label{Eq:SINR with edge user1}
&\SINR_{nk}^{\rm (Intra )}\left(\{\wb_{mi}\},\{\fb_{mj}\},\{\hb_{mnk}\}_{m=1}^{N_c}\right) \notag \\
&~~~~~~~~~~~~~~~~~~~~~=\frac{{\left|\hb_{nnk}^H\wb_{nk}\right|}^2}%
{\sum\limits_{i\neq k}^K {\left|\hb_{nnk}^H\wb_{ni}\right|}^2%
+\sum\limits_{m\neq n}^{N_c}\sum\limits_{i=1}^K {\left|\hb_{mnk}^H\wb_{mi}\right|}^2+ \sum\limits_{m=1 }^{N_c}\sum\limits_{j=1}^L|\hb_{mnk}^H\fb_{mj}|^2
+\sigma_{nk}^2},
\\
&\SINR_{\ell}^{\rm (Edge)}\left(\{\wb_{mi}\},\{\fb_{mj}\},\{\gb_{m\ell}\}_{m=1}^{N_c}\right) =\frac{\sum\limits_{m=1}^{N_c}{\left|\gb_{m\ell}^H\fb_{m\ell}\right|}^2}%
{\sum\limits_{m=1}^{N_c}\sum\limits_{i=1}^K {\left|\gb_{m\ell}^H\wb_{mi}\right|}^2%
+\sum\limits_{j\neq \ell}^{L}\sum\limits_{m=1}^{N_c} {\left|\gb_{m\ell}^H\fb_{mj}\right|}^2
+\sigma_{\ell}^2}. \label{Eq:SINR with edge user2}
\end{align}

Our goal here is, again, to find the beamforming vectors that are robust against the possible CSI errors. {As the channel error model for intra-cell MSs,
we model the cell-edge MSs' channel  as}
\begin{align}\label{CSIadditiveError2}
\gb_{m\ell}&=\ghb_{m\ell}+\vb_{m\ell}~~\forall~m\in\Nset,~\ell\in\Lset,
\end{align}
where $\ghb_{m\ell}\in \Cset^{N_t}$ is the preassumed CSI, and $\vb_{m\ell} \in \Cset^{N_t}$ is the CSI error
satisfying $\vb_{m\ell}^H\tilde\Qb_{m\ell}\vb_{m\ell}\leq 1$
in which $\tilde\Qb_{m\ell}\succ\zerob$.
We consider the following worst-case robust formulation:
\begin{subequations}\label{RPM edgeuser}
\begin{align}
  \min_{\substack{\{\wb_{nk}\},\{\fb_{n\ell}\}}}~&\sum_{n=1}^{N_c}{\alpha_n}
  \left(\sum_{k=1}^K{\left\|\wb_{nk}\right\|^2}+\sum_{\ell=1}^L{\left\|\fb_{n\ell}\right\|^2}\right)\\
  \st~~&
  \SINR_{nk}^{\rm (Intra )}\left(\{\wb_{mi}\},\{\fb_{mj}\},\{\hhb_{mnk}+\eb_{mnk}\}_{m=1}^{N_c}\right)\geq \gamma_{nk}
  \notag\\
  &~~~~~~~~~~~~~~~~~~~~~~~\forall~\eb_{mnk}^H\Qb_{mnk}\eb_{mnk}\leq 1,~m\in\Nset,~n\in\Nset,~k\in\Kset, \label{RPM edgeuser C1}\\
  &\SINR_{\ell}^{\rm (Edge)}\left(\{\wb_{mi}\},\{\fb_{mj}\},\{\ghb_{m\ell}+\vb_{m\ell}\}_{m=1}^{N_c}\right)\geq \gamma_\ell \notag\\
  &~~~~~~~~~~~~~~~~~~~~~~~\forall~\vb_{m\ell}^H\tilde\Qb_{m\ell}\vb_{m\ell}\leq 1,~m\in\Nset,~\ell\in\Lset. \label{RPM edgeuser C2}
\end{align}\end{subequations}
The proposed method based on SDR and {S}-lemma in Section \ref{subsec: sdr and s-lemma} can be used to handle the above problem \eqref{RPM edgeuser} as well. Firstly, {replace} each $\wb_{nk}\wb_{nk}^H$ and each $\fb_{n\ell}\fb_{n\ell}^H$ by general-rank $\Wb_{nk}\succeq \zerob$ and $\Fb_{n\ell}\succeq \zerob$, respectively. Secondly, {follow} the steps as in \eqref{E0040} to \eqref{LMI2} to decouple and transform the worst-case constraints in \eqref{RPM edgeuser C1} and \eqref{RPM edgeuser C2} into a finite number of LMIs. The obtained SDR problem can be shown {to be} the following SDP:
\vspace{0.3cm}
\begin{align}\label{RobustMCBF edgeuser}
\min_{\substack{\{\Wb_{nk}\},\\\{\lambda_{mnk}\},\{t_{mnk}\}\\ \{\Fb_{nk}\},\\\{\tilde\lambda_{m\ell}\},\{\eta_{m\ell}\}}}~
  &\sum_{n=1}^{N_c}\alpha_n \left(\sum_{k=1}^K\Tr(\Wb_{nk})+\sum_{\ell=1}^L\Tr(\Fb_{n\ell})\right)\\
  \st~
  &\Phib_{nk} \left(\{\Wb_{ni}\}_{i=1}^K,\{t_{mnk}\}_{m},\lambda_{nnk}\right)
   -\begin{bmatrix} \Ib_{N_t} \\  \hhb_{nnk}^H  \end{bmatrix}
    \left(\sum_{j=1}^L \Fb_{nj}\right)
    \begin{bmatrix} \Ib_{N_t} \\  \hhb_{nnk}^H  \end{bmatrix}^H
   \succeq \zerob,\notag \\
  &\Psib_{mnk}\left(\{\Wb_{mi}\}_{i=1}^K,  t_{mnk},            \lambda_{mnk}\right)
  -\begin{bmatrix} \Ib_{N_t} \\  \hhb_{mnk}^H  \end{bmatrix}
    \left(\sum_{j=1}^L \Fb_{mj}\right)
    \begin{bmatrix} \Ib_{N_t} \\  \hhb_{mnk}^H  \end{bmatrix}^H
    \succeq \zerob
  ~\forall~ m\neq n,  \notag \\
  &\Wb_{nk}\succeq \zerob,~\lambda_{mnk}\geq 0~\forall~m,n\in\Nset,~k\in \Kset\notag,\\
  &\sum_{m=1}^{N_c} \eta_{m\ell}-\sigma_\ell^2 \geq 0, \notag \\
  &\begin{bmatrix} \Ib_{N_t} \\  \ghb_{m\ell}^H  \end{bmatrix}
    \left(\frac{1}{\gamma_{\ell}}\Fb_{m\ell}\!-\!\sum_{i=1}^K \Wb_{mi}-\sum_{j\neq \ell}^L \Fb_{mj}\right)
    \begin{bmatrix} \Ib_{N_t} \\  \ghb_{m\ell}^H  \end{bmatrix}^H
   +\begin{bmatrix} \tilde\lambda_{m\ell}\tilde\Qb_{m\ell}&\zerob\\\zerob & -\!\tilde\lambda_{m\ell}\!\!-\!\eta_{m\ell}
   \end{bmatrix}\succeq \zerob, \notag \\
   &\Fb_{m\ell}\succeq \zerob,~\tilde\lambda_{m\ell}\geq 0~\forall~ m\in\Nset,~\ell\in\Lset, \notag
\end{align}
where $\{\tilde\lambda_{m\ell}\},$ $\{\eta_{m\ell}\}$ are slack variables. For the spherical error model, a sufficient condition for the tightness of SDR, which is similar to Proposition \ref{prop: rank-one}, can be {shown to}
\begin{align}\label{condition}
   &\varepsilon_{nmk}\leq \bar\varepsilon_{nmk},~\varepsilon_{n\ell}\leq \bar\varepsilon_{n\ell} \notag\\
   &
   \varepsilon_{nnk}   < \sqrt{\frac{\alpha_n\gamma_{nk}\sigma_{nk}^2}{g^\star}},~\text{and}~
   \varepsilon_{n\ell} < \sqrt{\frac{\alpha_n\gamma_{\ell}\sigma_{\ell}^2}{g^\star} } ~\forall~n,m,k,\ell,
\end{align}
where $\{\bar\varepsilon_{nmk}\}$, $\{\bar\varepsilon_{n\ell}\}$ are the CSI error bounds for which problem \eqref{RobustMCBF edgeuser} is feasible, and $g^\star>0$ is the associated optimal objective value. The condition in \eqref{condition} implies that problem \eqref{RobustMCBF edgeuser} can attain the global optimum of \eqref{RPM edgeuser} if the CSI errors are sufficiently small.

A distributed optimization algorithm for problem \eqref{RobustMCBF edgeuser} can also be developed by applying ADMM, using the same ideas as presented in Section \ref{subsec: distributed algorithm} for problem \eqref{RobustMCBF}. We will provide a simulation example in Section \ref{Sec:Simulation} to demonstrate the {efficacy of formulation \eqref{RobustMCBF edgeuser} in providing guaranteed QoS for the cell-edge MSs.} 

\section{Simulation results}\label{Sec:Simulation}
In this section, some simulation results are presented to examine the performance of the proposed
robust MCBF design and the distributed optimization algorithm (Algorithm 2).
The performance of the robust fully coordinated BF design in the previous section will also be
examined.

\subsection{Simulation Setting}\label{subsec:Simulation setting}


In the simulations, we not only consider the small scale channel fading but also the large scale fading effects such as
shadowing and {path} loss, in order to simulate the multi-cell scenario. Specifically, we follow the channel model
\cite{3GPP_standard2,Dahrouj10TWC}:
\begin{align}
\hb_{mnk}=10^{{-(128.1+37.6{\log_{10}}(d_{mnk}))}/{20}}\cdot\psi_{mnk}\cdot\varphi_{mnk}\cdot
(\hhb_{mnk}+\eb_{mnk}),\label{channel model}
\end{align}
where the exponential factor is due to the path loss depending on the distance between the $m$th BS and $\MSnk$ (denoted by $d_{mnk}$ in kilometers), $\psi_{mnk}$ reflects the shadowing effect, and $\varphi_{mnk}$ represents the transmit-receive antenna gain. The term inside the {parentheses} in \eqref{channel model} denotes the small scale fading which consists of the preassumed CSI $\hhb_{mnk}$ and the CSI error $\eb_{mnk}$. As seen from \eqref{channel model}, it is assumed that the BSs can accurately track the large scale fading, and suffers only from the small scale CSI errors.

The inter-BS distance is 500 meters, and the locations of the MSs in each cell are randomly determined with the distance to the serving BS at least 35 meters, i.e., $d_{nnk}\geq 0.035$ for all $n,k$.
The shadowing coefficient $\psi_{mnk}$ follows the log-normal distribution with zero mean and standard deviation equal to 8. The elements of the preassumed CSI $\{\hhb_{mnk}\}$ are i.i.d. complex Gaussian random variables with zero mean and unit variance. We also assume that all MSs have the same noise power spectral density equal to -162 dBm/Hz (-92 dBm {over a} 10 MHz bandwidth), and each BS has a maximum power limit 46 dBm \cite{3GPP_standard2}.
The SINR requirements of MSs are set the same, i.e., $\gamma_{mnk}\triangleq \gamma$, and each link has the same antenna gain $\varphi_{mnk}=15$ dBi. The power weight $\alpha_n$ for $\BSn$ is set to one for all $n$ (i.e., sum power). For the CSI errors, the spherical error model is considered, i.e., $\Qb_{mnk}=(1/\varepsilon_{mnk}^2)\Ib_{N_t}$ for all $m$, $n$ and $k$.
If not mentioned specifically, the error radii $\varepsilon_{mnk}$ are set the same and denoted by $\varepsilon$.

\subsection{Performance Comparison with Existing Methods}

For the robust MCBF design \eqref{RPM}, we first compare the proposed SDR method with the convex restrictive approximation method
in \cite{Tajer11TSP}. The single-cell beamforming (SCBF) design with independent ICI constraints \cite{Huh2010,Shen2011} is
also compared. All the design formulations are solved by \texttt{SeDuMi} \cite{SeDuMi}.

\begin{figure*}[t]\centering%
{\psfrag{Tajer}[c][c][0.7]{\cite{Tajer11TSP})}
\subfigure[{$N_c=2$}] {\includegraphics[width=0.47\linewidth]{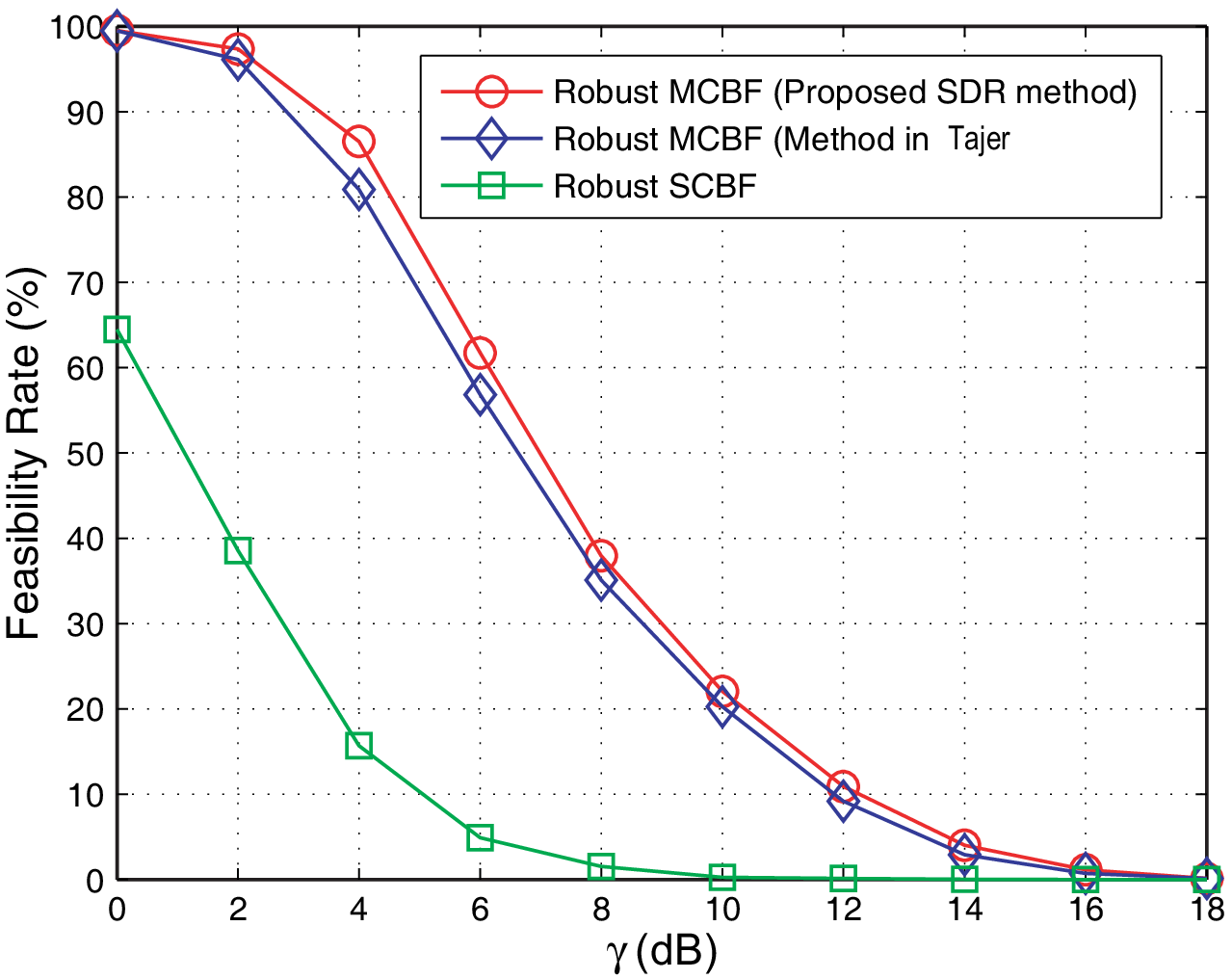}}}~%
\hspace{0.3cm}
{\psfrag{Tajer}[c][c][0.7]{\cite{Tajer11TSP})}
\subfigure[{$N_c=3$}] {\includegraphics[width=0.47\linewidth]{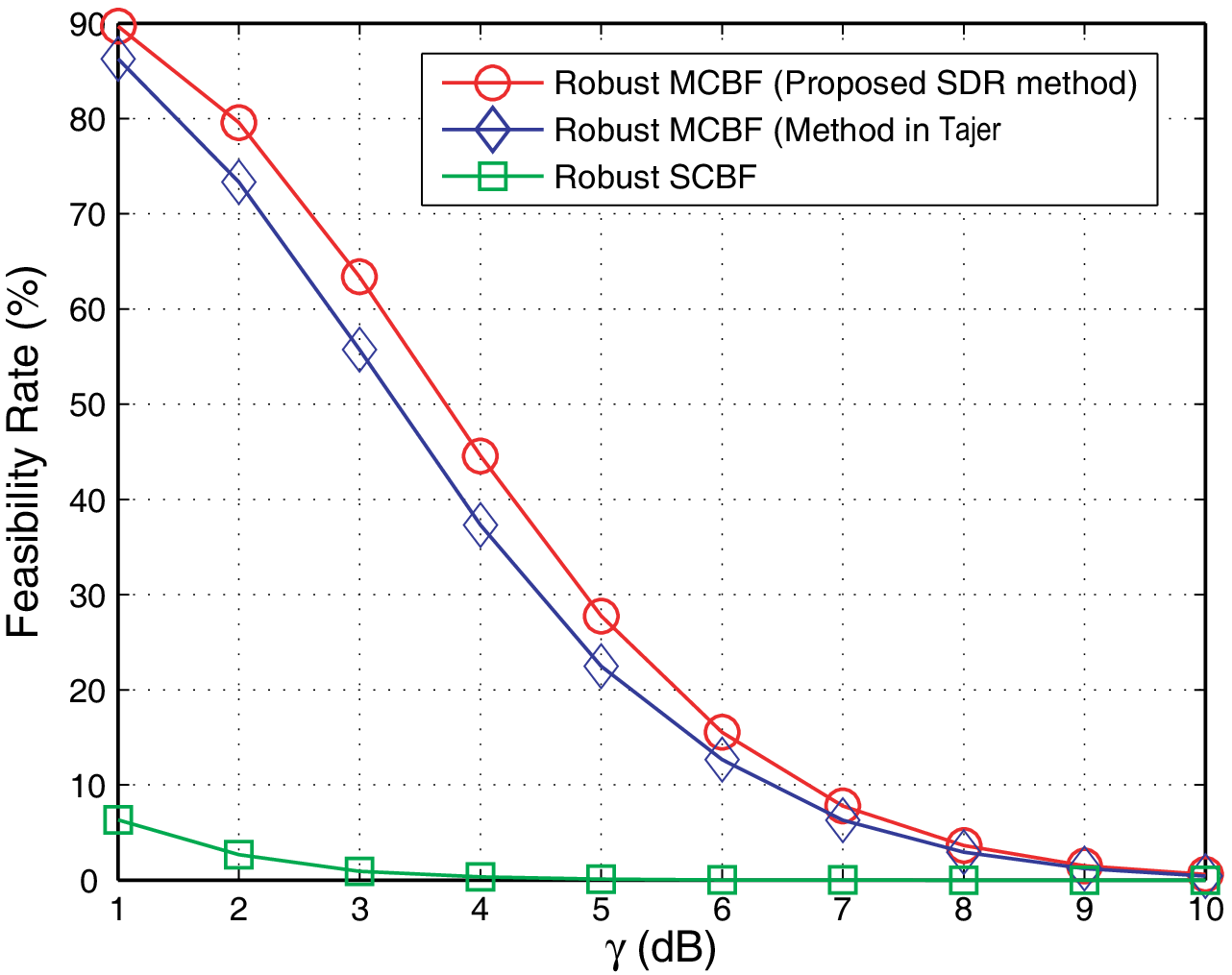}}}%
\vspace{-0.3cm}\caption{Feasibility rate (\%) versus SINR requirement $\gamma$ for
$K=4$, $N_t=6$, $\varepsilon=0.1.$
}
\label{Fig:MCBFfeas}\vspace{-0.6cm}
\end{figure*}

We first present the feasibility rates of the three beamforming designs. We say that the formulation under test is
feasible if it can yield an optimal solution with each BS's power no greater than 46 dBm. Figure \ref{Fig:MCBFfeas} presents the simulation results for $K=4$ (MSs/cell), $N_t=6$ (antennas/BS), and (a) $N_c=2$ and (b) $N_c=3$, respectively. The CSI error radius $\varepsilon$ is set to $0.1$.
Seven thousand channel realizations are tested. We can observe from this figure that the robust MCBF design
exhibits a much higher feasibility rate than the SCBF design, showing the improved capability of coordinated beamforming by
exploiting the degrees of freedom provided by multiple BSs. Secondly, we can see that the robust MCBF design using the proposed SDR method exhibits a slightly higher feasibility rate than that using the method in \cite{Tajer11TSP}. We should emphasize that in the simulation tests, the SDR problem \eqref{intraRPM SDR} all yields rank-one solutions.
Hence, the feasibility rate of the SDR method is in fact that of the original problem \eqref{RPM}.

%

\begin{figure*}[t]\centering%
{\psfrag{Tajer}[c][c][0.7]{\cite{Tajer11TSP})}
\subfigure[{$\varepsilon=0.1$}]
{\includegraphics[width=0.47\linewidth]{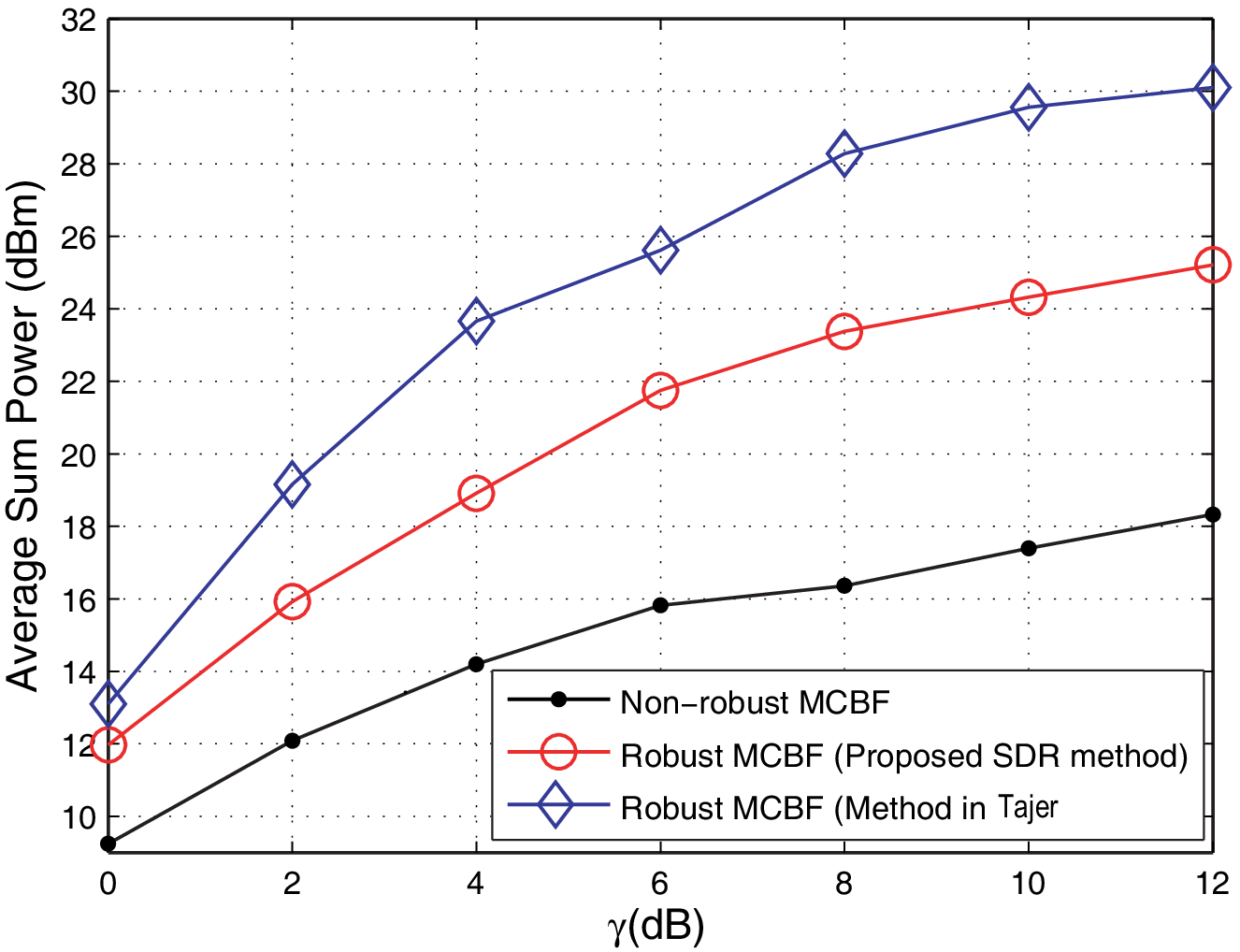}}}~%
\hspace{0.3cm} {\psfrag{Tajer}[c][c][0.7]{\cite{Tajer11TSP})}
\subfigure[{$\gamma=10$ dB}]
{\includegraphics[width=0.47\linewidth]{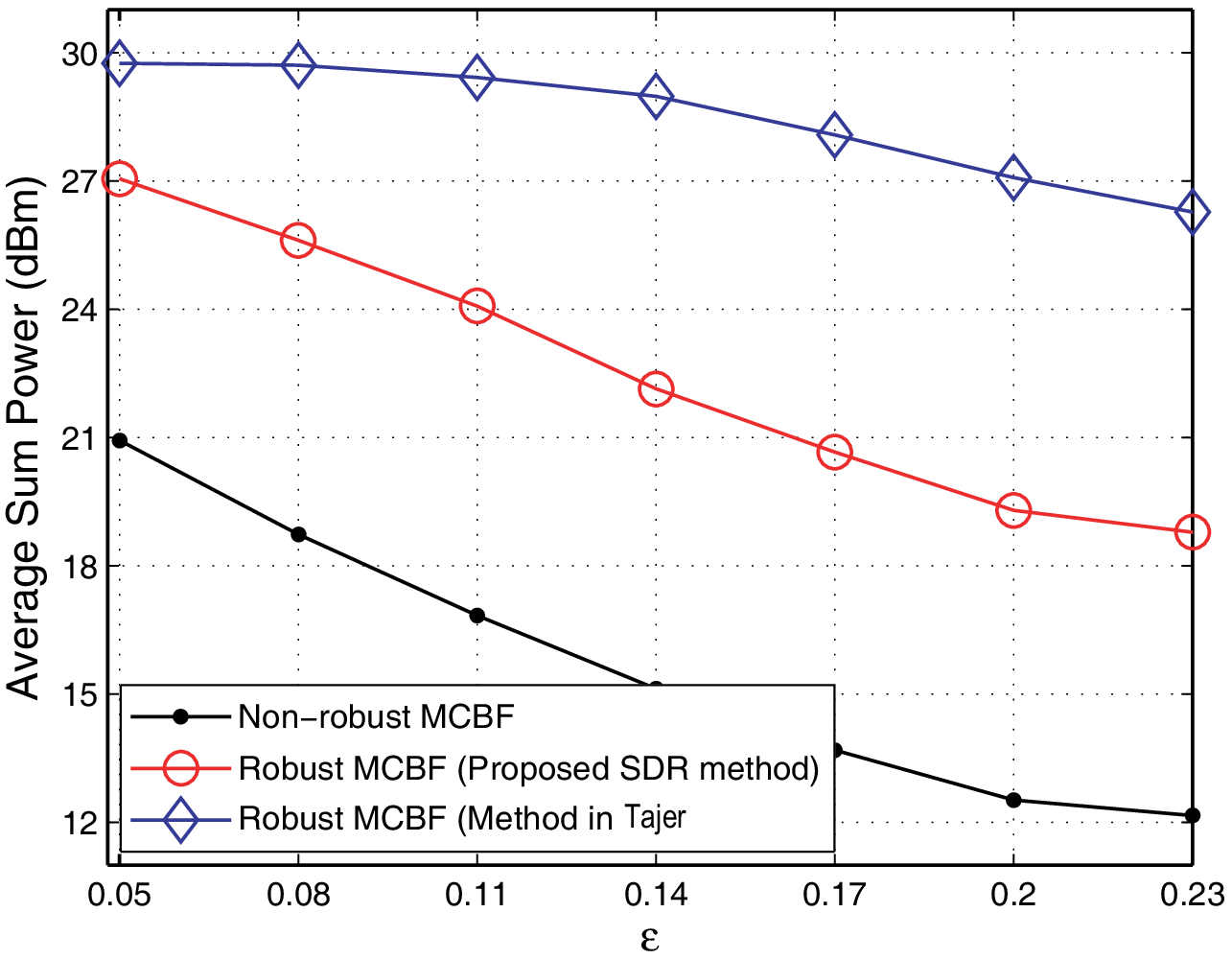}}}
\vspace{-0.3cm}\caption{Average transmission sum power (dBm) of various methods for $N_c=2$, $K=4$ and $N_t=6$.}
\label{Fig:MCBFpower}\vspace{-0.5cm}
\end{figure*}

Next, we examine the average transmission sum powers of the robust MCBF design. As a performance benchmark, we also present the average sum power of the non-robust MCBF design \eqref{PM}. Figure \ref{Fig:MCBFpower}(a) shows the results of average sum power (dBm) versus the SINR requirement $\gamma$ for $N_c=2$, $K=4$, $N_t=6$, and $\varepsilon=0.1$.
7,000 channel realizations are tested and each of the results in Figure \ref{Fig:MCBFpower}(a) is obtained by averaging over the feasible channel realizations at each $\gamma$. One can observe from the figure that, as a price for worst-case performance guarantee, the robust MCBF designs require higher average transmission powers than the non-robust design. Comparing the proposed SDR method with the method in \cite{Tajer11TSP}, one can see that the proposed SDR method is more power efficient. For example, for $\gamma=10$ dB, the proposed SDR method consumes around 24 dBm while the method in \cite{Tajer11TSP} requires 29 dBm. Figure \ref{Fig:MCBFpower}(b) displays the results of average transmission sum power (dBm) versus the CSI error radius $\varepsilon$, for $\gamma=10$ dB. As seen, the proposed SDR method is much more power
efficient than the method in \cite{Tajer11TSP}.

\vspace{-0.3cm}
\subsection{Performance of Proposed Distributed Robust MCBF Algorithm}
\begin{figure*}[!t]\centering%
  {\subfigure[{$N_c=2, K=2, N_t=8$}]
  {\includegraphics[width=0.95\linewidth]{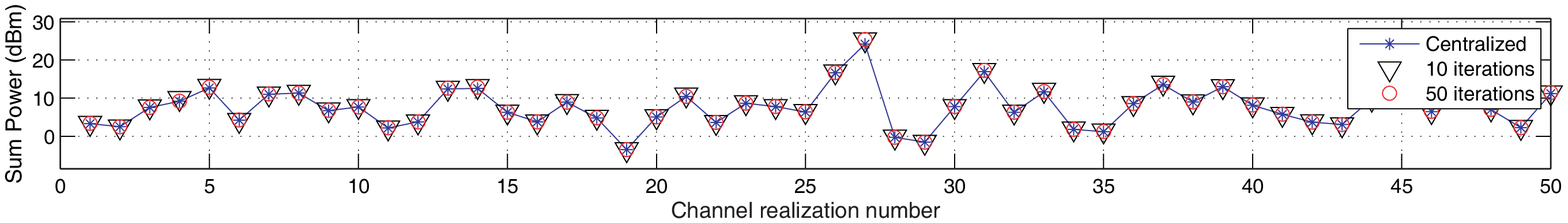}}}\\\vspace{0.1cm}%
  {\subfigure[{$N_c=2, K=4, N_t=8$}]
  {\includegraphics[width=0.95\linewidth]{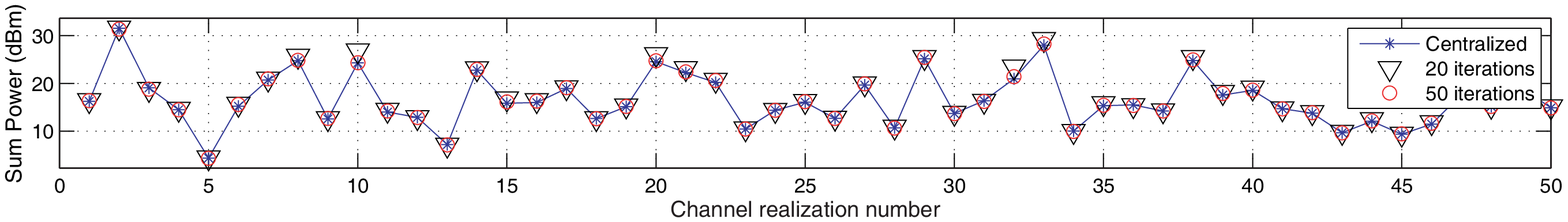}}}\\\vspace{0.1cm}%
  {\subfigure[{$N_c=3, K=3, N_t=8$}]
  {\includegraphics[width=0.95\linewidth]{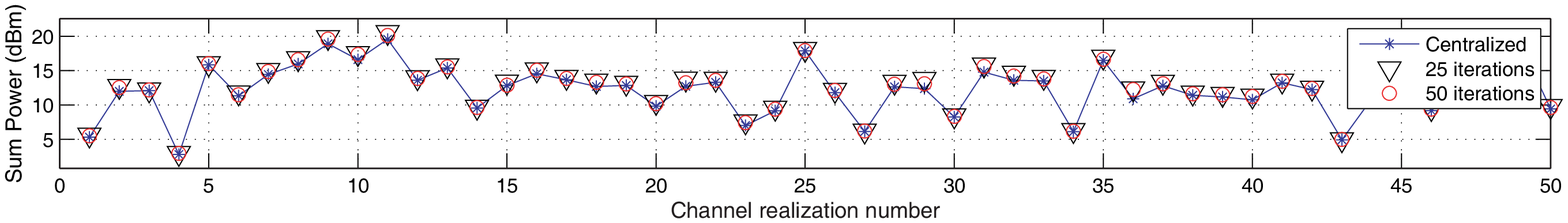}}}\\\vspace{0.1cm}%
  {\subfigure[{$N_c=8, K=1, N_t=4$}]
  {\includegraphics[width=0.95\linewidth]{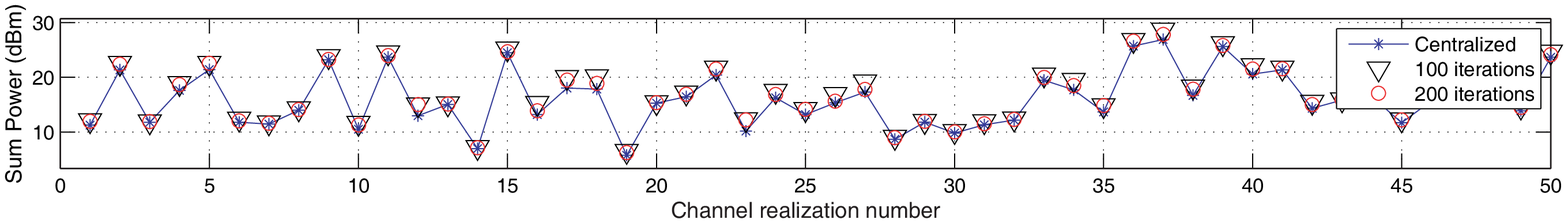}}}\vspace{-0.1cm}%
  \vspace{-0.3cm}\caption{Sum power comparison between the centralized robust MCBF solution and the distributed robust MCBF solution (by Algorithm 2) {over} 50 randomly generated channel realizations, {for} $\gamma=10$ dB and $\varepsilon=0.05$.}
  \label{Fig:50channels}\vspace{-0.9cm}
\end{figure*}

Now, let us examine the performance of the proposed distributed robust MCBF algorithm (Algorithm \ref{RobustDMCBF}). In the simulations, the initial input values $\{\nub_n(0),\mu_n(0),\tb(0),\rho_n(0)\}_{n=1}^{N_c}$ are all set to zero. The augmented penalty parameter $c$ is not fixed as a constant; instead it follows the rule
\begin{align}\label{c update}
    c(q+1):= \left\{\vphantom{\begin{array}{c} a\\[2ex]\end{array}}\right.
    \kern-7pt
    \begin{array}{ll} qc(q) & {\rm if~}c(q)<1, \\
                          1 & {\rm otherwise},\\
   \end{array}
\end{align}
where $c(0)=10^{-6}$.
As one can verify that $c(q)$ will reach the value of one after nine iterations, Algorithm \ref{RobustDMCBF} following \eqref{c update} will still converge to the global optimum for a sufficiently large $q$, according to Proposition \ref{prop: convergence}.
Firstly, we compare the optimal sum power of the centralized problem \eqref{RPM} with that obtained by Algorithm 2. The simulation results by testing over 50 randomly generated channel realizations are presented in Fig. \ref{Fig:50channels}, under various simulation settings.
From Fig. \ref{Fig:50channels}(a) {and} Fig. \ref{Fig:50channels}(b), where $N_c=2$ and $K=2$ and $K=4$, respectively, we can observe that Algorithm 2 can yield near-optimal solutions within 50 iterations.
As observed, for most of the cases, 10 and 20 iterations are quite sufficient for the scenarios in Fig. \ref{Fig:50channels}(a) and Fig. \ref{Fig:50channels}(b), respectively. When the the number of cells $N_c$ increases to three ($N_c=3$), as shown in Fig. \ref{Fig:50channels}(c), 25 iterations are sufficient to obtain a near-optimal solution. General speaking, as the number of cells and {that of} MSs increase, the number of iterations needed to achieve a near-optimal performance also increases.
As seen from Fig. \ref{Fig:50channels}(d), where $N_c=8$, at leat 100 iterations are required.

\begin{figure*}[!h]\centering%
{\subfigure[$N_t= 8, \gamma=10$ dB, $\varepsilon=0.05$]
{\includegraphics[width=.5\linewidth]{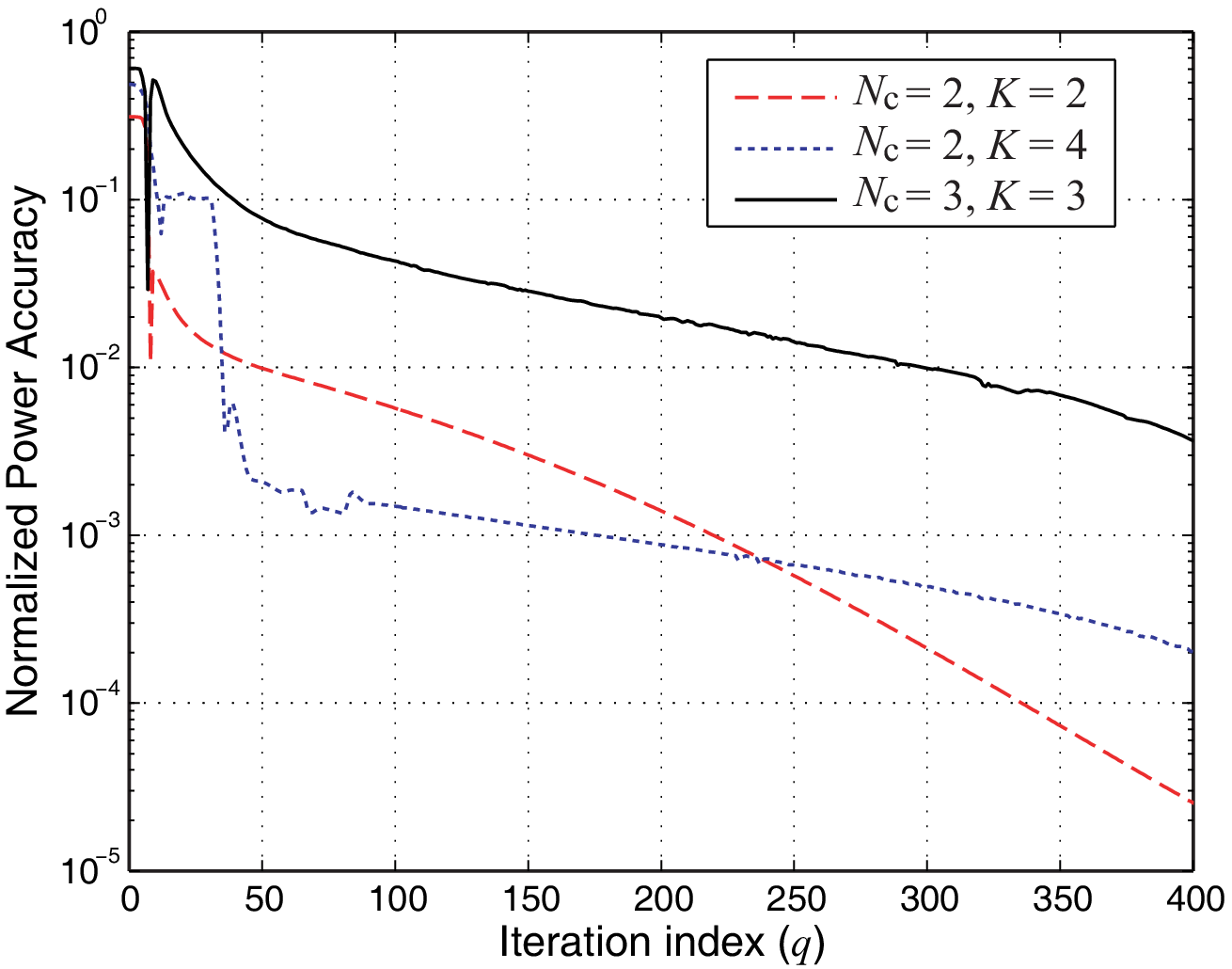}}}~%
{\subfigure[$N_c=2, K=2, N_t=4$]
{\includegraphics[width=.52\linewidth]{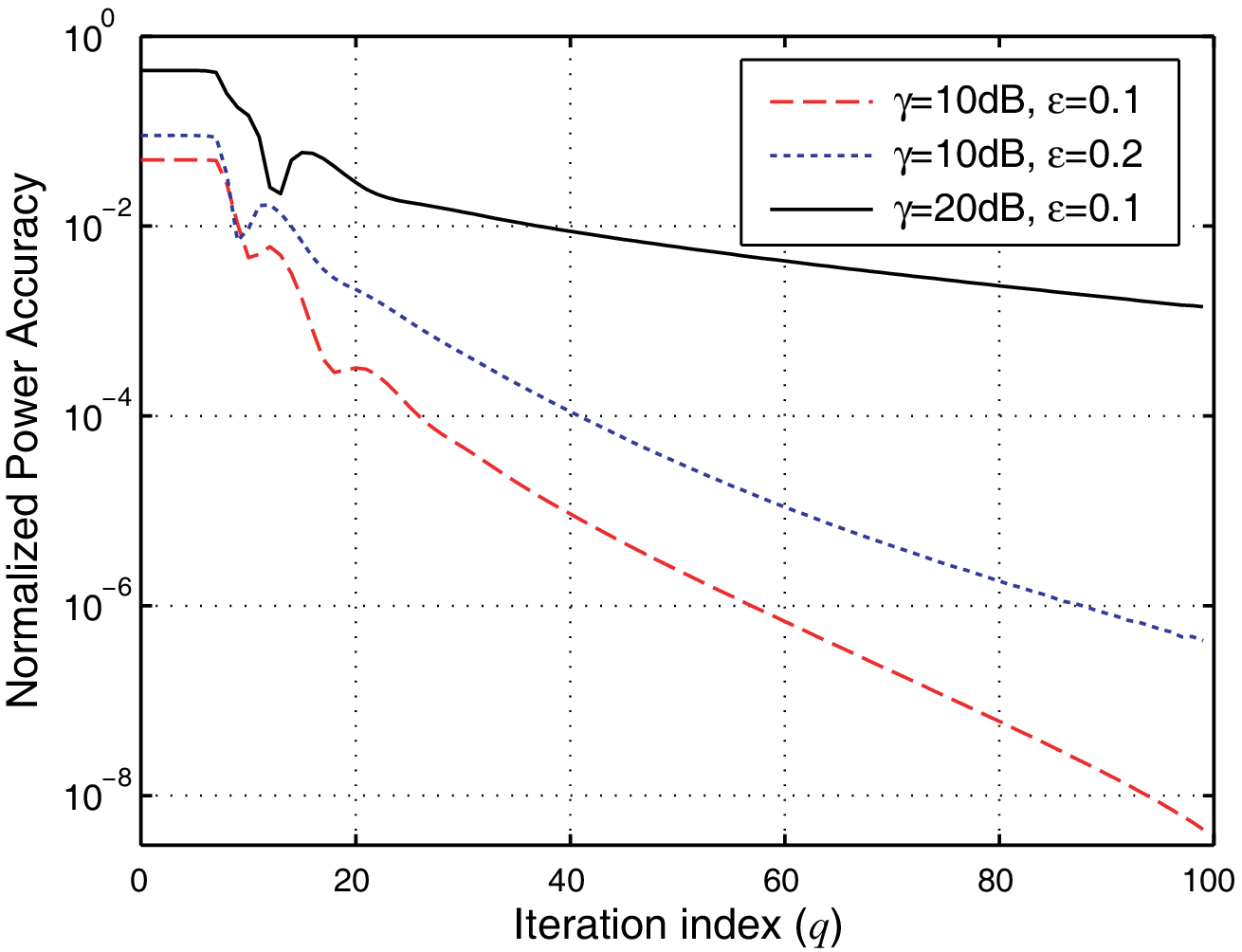}}}\\%
\vspace{-0.3cm}\caption{Typical convergence curves of Algorithm 2 under various simulation settings.}
\label{Fig:convergence 2}\vspace{-0.5cm}
\end{figure*}

To further look into the convergence behavior of Algorithm 2, we show in Fig. \ref{Fig:convergence 2}(a) the typical convergence curves of Algorithm 2 in the scenarios considered in Fig. \ref{Fig:50channels}(a) to Fig. \ref{Fig:50channels}(c). In Fig. \ref{Fig:convergence 2}, the normalized power accuracy is defined as
\begin{align}
  \text{Normalized power accuracy}=\frac{|P^{\star}(q)-P^\star|}{P^\star},
\end{align}
where $P^{\star}(q)=\sum_{n=1}^{N_c}p_i(q)$ is the sum power at iteration $q$, and $P^\star$ denotes the centralized solution of problem \eqref{RobustMCBF}.
We can see from Fig. \ref{Fig:convergence 2}(a) that Algorithm 2 can yield a solution with the normalized power accuracy {smaller} than $0.1$ within 50 iterations.
Figure \ref{Fig:convergence 2}(b) presents the convergence curves of Algorithm 2 for $N_c=2$, $K=2$, $N_t=4$, and various CSI error radii and SINR requirements.
It can be seen from this figure that the convergence speed can be slowed down as the CSI error radius or SINR requirement increases. Nevertheless, for the scenarios considered in Fig. \ref{Fig:convergence 2}(b), less than 40 iterations are needed for achieving 0.01 normalized power accuracy. The simulation results presented in Fig. \ref{Fig:50channels} and Fig.\ref{Fig:convergence 2} well demonstrate the convergence of Algorithm 2, as stated in Proposition \ref{prop: convergence}.

\vspace{-0.5cm}
\subsection{Performance of Robust Fully Coordinated BF}

\begin{figure*}[!t]\centering{
  \psfrag{BS1}[c][c][1]{BS$_1$}
  \psfrag{BS2}[c][c][1]{BS$_2$}
  \psfrag{BS3}[c][c][1]{BS$_3$}
\includegraphics[width=0.44\linewidth]{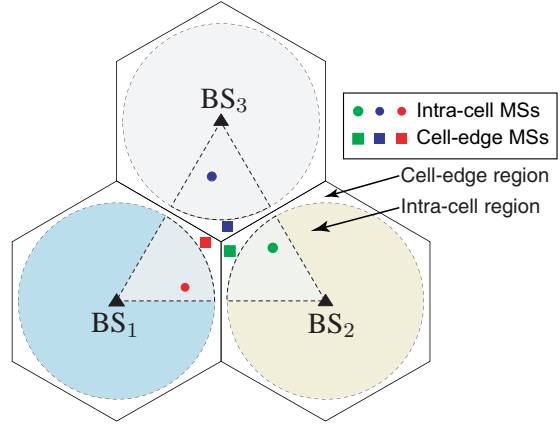}}\vspace{0.1cm}
\caption{Illustration of the simulation scenario for the robust fully coordinated BF design \eqref{RPM edgeuser}.
}\label{Fig:3cellmodel}\vspace{-0.1cm}
\end{figure*}

\begin{figure*}[!t]\centering%
\psfrag{SINR (dB)}[c][c][0.9]{$\gamma$ (dB)}
{\subfigure[]{\includegraphics[width=.47\linewidth]
{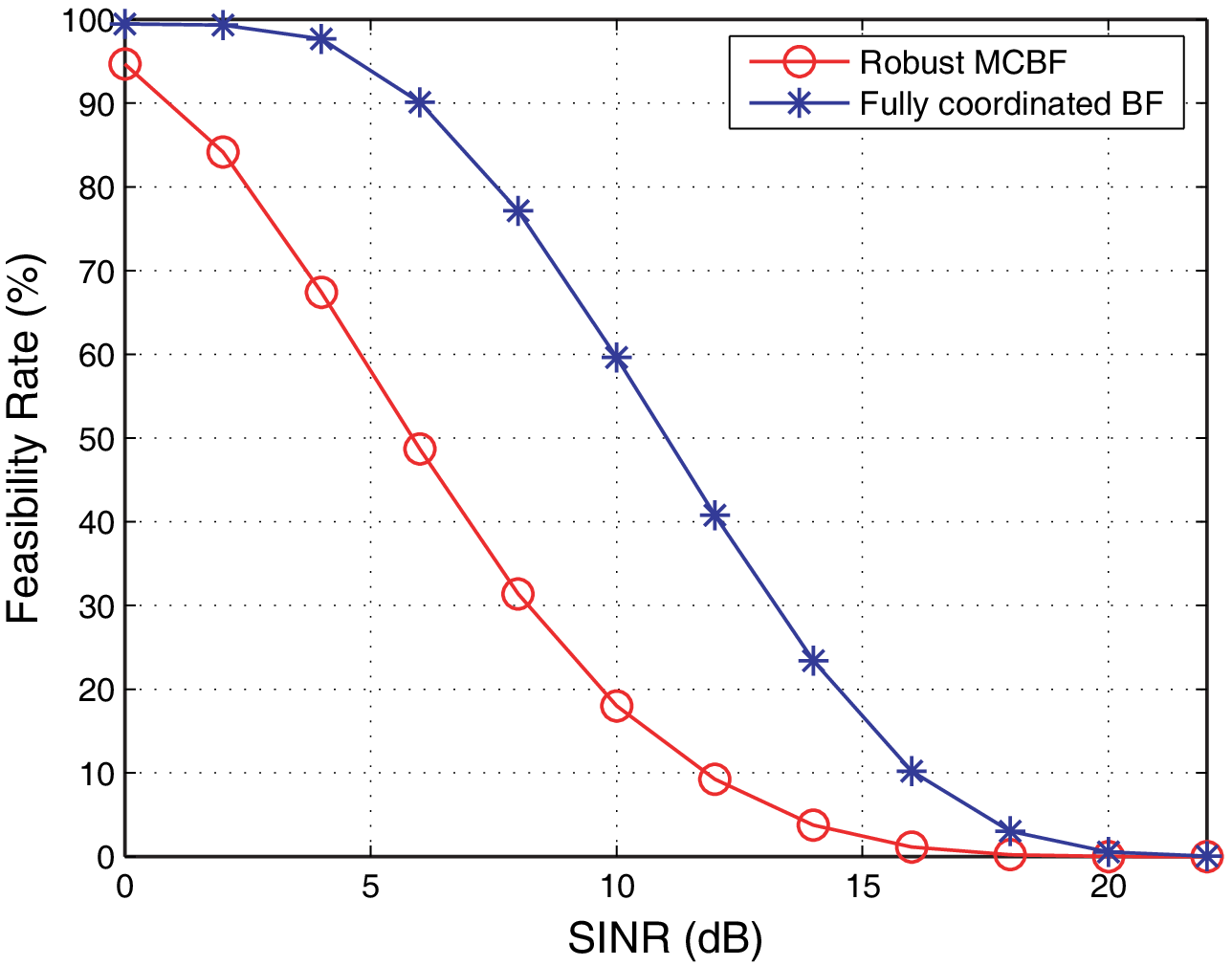}}}~\hspace{0.3cm}%
{\subfigure[]{\includegraphics[width=.47\linewidth]
{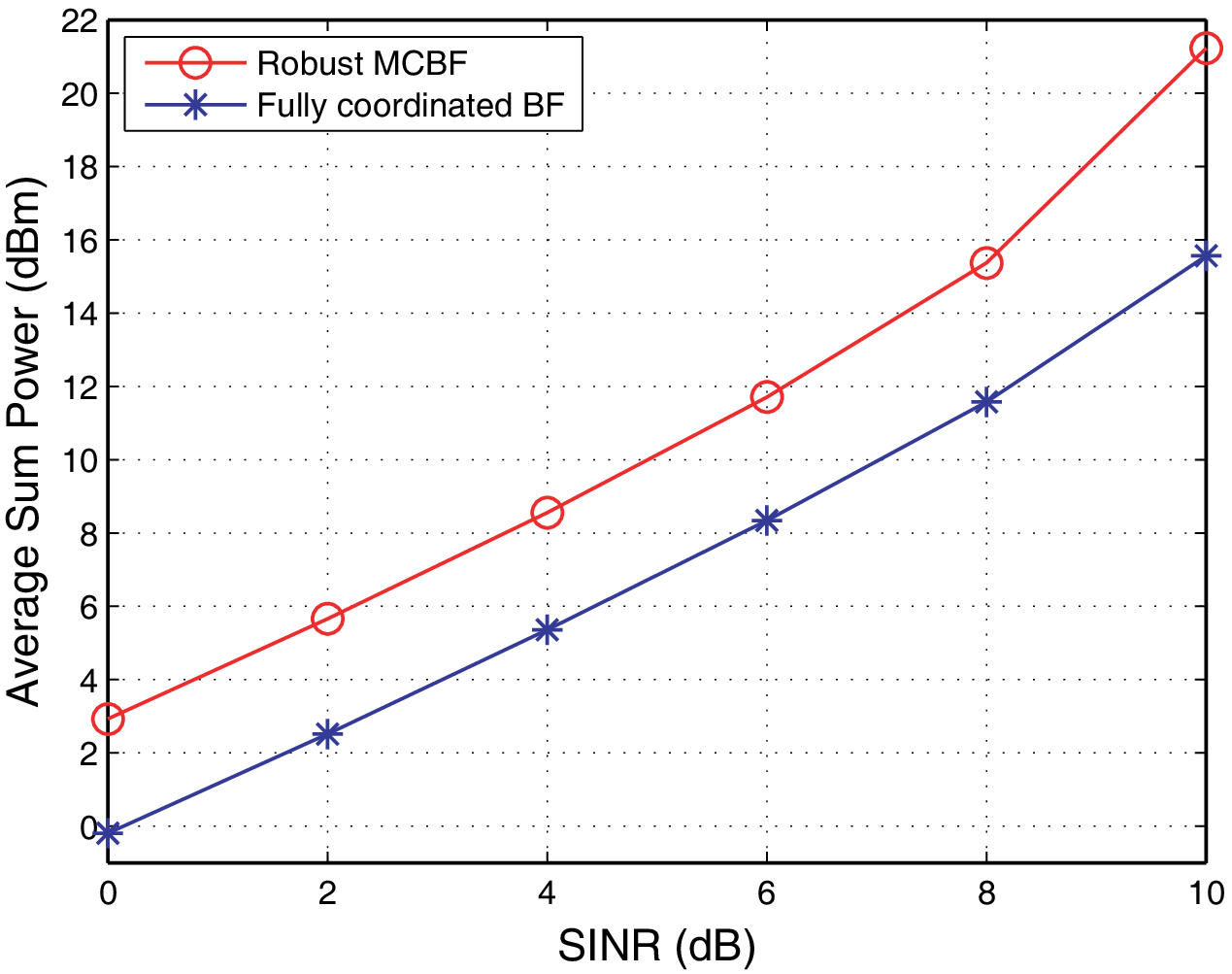}}}\\\vspace{-0.4cm}%
\caption{{Performance comparison results of the robust MCBF design \eqref{RPM} and the fully coordinated BF design \eqref{RPM edgeuser}}, for $N_c=3$, $K=2$, $N_t=4$, and $\varepsilon=0.1$.}
\label{Fig:FullyCBF}\vspace{-0.9cm}
\end{figure*}

In this subsection, we examine the effectiveness of the robust fully coordinated BF design \eqref{RPM edgeuser} in serving the cell-edge MSs. To this end, let us consider a three-cell system ($N_c=3$) with two MSs in each cell ($K=2$), as illustrated in Fig. \ref{Fig:3cellmodel}.
We divide each cell into two parts, namely, the intra-cell region and the cell-edge region. In particular, the inter-BS distance is set to 500 meters and the radius for the intra-cell region is 235 meters. In each cell, the position of one of the MSs is randomly chosen within the intra-cell region; while the other MS is randomly located in the cell-edge region within the equilateral triangle formed by the three BSs (see Fig. \ref{Fig:3cellmodel}). As the robust fully coordinated BF design \eqref{RPM edgeuser} is applied, the three MSs in the cell-edge regions will be served simultaneously by the three BSs, i.e., $K=1$ and $L=3$ in \eqref{RPM edgeuser}.
Figure \ref{Fig:FullyCBF} shows the performance comparison results of the robust fully coordinated BF design \eqref{RPM edgeuser} and the robust MCBF design \eqref{RPM} by testing over 17,000 channel realizations. The SDR formulation \eqref{RobustMCBF edgeuser} is used as an approximation to \eqref{RPM edgeuser}. It is found in this simulation test that SDR formulation \eqref{RobustMCBF edgeuser} always yields rank-one solutions; hence the obtained solution is exactly the optimal solution of \eqref{RPM edgeuser} for the tested problem instances.
From Fig. \ref{Fig:FullyCBF}, we can observe that the robust fully coordinated BF design is more feasible and is more power efficient (for around 3 dB) than the robust MCBF design in serving cell-edge MSs.

\section{Conclusions}\label{sec:conclusions}
In this paper, we have investigated the worst-case SINR constrained robust MCBF design problem [in \eqref{RPM}]. While {the robust design problem} involves complicated nonconvex worst-case SINR constraints, we have presented an efficient approximation method based on SDR. We have shown that when there is only one MS in each cell or when the CSI errors are sufficiently small, the proposed SDR method can yield the global optimal solution to the original problem (Proposition \ref{prop: rank-one}).
Moreover, by using ADMM, we have proposed a distributed robust MCBF algorithm (Algorithm 2). {The proposed distributed algorithm is appealing because it is proven to converge to the global optimum of the centralized problem, with a much less backhaul signaling overhead compared to the existing methods (Proposition \ref{prop: convergence})}.
Extension of the proposed {SDR method} to a fully coordinated scenario has been presented as well. {The presented simulation results have shown} that the proposed SDR method is more power efficient than the existing method, and {that} the proposed distributed optimization algorithm can obtain beamforming solutions with the normalized power accuracy {smaller} than $0.1$ in {twenties of iterations} for the typical scenario of $N_c\leq 3$.
The robust fully coordinated BF design \eqref{RPM edgeuser} has also been shown more power efficient if there are MSs located in the cell boundary.

\appendices {\setcounter{equation}{0}
\renewcommand{\theequation}{A.\arabic{equation}}

\section{Proof of Proposition \ref{prop: rank-one}}\label{proof of rank one}

\emph{Proof of case C1):} We rewrite problem \eqref{RobustMCBF} for $K=1$ as follows
\begin{subequations}\label{RobustMCBF_K1}
\begin{align}
\min_{\substack{\{\Wb_{n}\},\\\{\lambda_{mn}\},\{t_{mn}\}_{m\neq
n}}}~
  &\sum_{n=1}^{N_c}\alpha_n \Tr(\Wb_{n})\\
  \st~
  &\Phib_{n} \left(\Wb_{n},\{t_{mn}\}_{m\neq n},\lambda_{nn}\right) \succeq \zerob,\label{RobustMCBF-b}\\
  &\Psib_{mn}\left(\Wb_{m},  t_{mn},            \lambda_{mn}\right) \succeq \zerob
  ~\forall~ m\neq n,  \label{RobustMCBF-c}\\
  &\Wb_{n}\succeq \zerob,~\lambda_{mn}\geq 0~\forall~m,n\in\Nset,\label{RobustMCBF-e}
\end{align}
\end{subequations}
where the subindices $k$ and $i$ are removed for notational simplicity. Proposition \ref{prop: rank-one} can be proved by investigating the KKT conditions of problem \eqref{RobustMCBF_K1}.
According to the KKT conditions, one can verify that the optimal
$\{\Wb_{n}^\star\}$, $\{\lambda_{mn}^\star\},\{t_{mn}^\star\}$ satisfy the following conditions
\begin{align}\label{C1}
   & \Zb_n^\star\Wb_n^\star=\zerob,~\Wb_n^\star\neq \zerob,\\
   & \Zb_n^\star=\Ib_{N_t} + \sum_{m\neq n}^{N_c}
        \begin{bmatrix}\Ib_{N_t} &  \hhb_{nm}  \end{bmatrix}  \Yb_{nm}^\star
        \begin{bmatrix}\Ib_{N_t} \\ \hhb_{nm}^H\end{bmatrix}
      -\frac{1}{\gamma_n}
        \begin{bmatrix}\Ib_{N_t} & \hhb_{nn}  \end{bmatrix}   \Yb_{nn}^\star
        \begin{bmatrix}\Ib_{N_t} \\\hhb_{nn}^H\end{bmatrix}
    \succeq \zerob, \label{C2}\\
   & \Phib_{n}\left(\Wb_{n}^\star,\{t_{mn}^\star\}_{m\neq n},\lambda_{nn}^\star\right)\Yb_{nn}^\star =\zerob,~
     \Yb_{nn}^\star\neq \zerob,\label{C3}\\
   & \Tr(\Qb_{nn}\Ab_n) \leq c_n,\label{C9}\\
   & t_{mn}^\star > 0~\forall~m\neq n, \label{C4} \\
   & \lambda_{nn}^\star > 0, \label{C5}
\end{align}
for $n=1,\ldots,N_c$, where $\Zb_n^\star \in \mathbb{H}^{N_t}$,
\begin{align}\label{part Y}
{\Yb}_{nn}^\star\triangleq\begin{bmatrix}\Ab_n&\bb_n\\\bb_n^H&c_n\end{bmatrix}
\succeq \zerob,
\end{align}
and $\Yb_{nm}^\star \succeq \zerob$, $m\neq n$, are the associated dual variables. Specifically, to show \eqref{C5}, we note that if
$\lambda_{nn}^\star =0$, {then it follows from \eqref{LMI1} that}
\begin{align}
  [-\hhb_{nn}^H~1] \Phib_{n} \left(\Wb_{n}^\star,\{t_{mn}^\star\}_{m\neq n},\lambda_{nn}^\star\right)
  \begin{bmatrix}-\hhb_{nn} \\ 1\end{bmatrix}=-\sum\limits_{m\neq n}t_{mn}-\sigma_{n}^2<0,
\end{align} which contradicts with \eqref{RobustMCBF-b}.

The first step of the proof is to show that $\Wb_{n}^\star$ has rank one whenever $\Yb_{nn}^\star$ has rank one. Suppose that
$\Yb_{nn}^\star=\yb\yb^H$ where $\yb \in \mathbb{C}^{N_t+1}$, i.e.,
$\Yb_{nn}^\star$ is of rank one. Let $\Xb_n\triangleq
\Xb_n^{1/2}\Xb_n^{1/2}=\Ib_{N_t} +\sum\nolimits_{m\neq n}^{N_c}
  \begin{bmatrix}\Ib_{N_t} & \hhb_{nm}\end{bmatrix}\Yb_{nm}^\star
  \begin{bmatrix}\Ib_{N_t}\\\hhb_{nm}^H\end{bmatrix}\succ \zerob$.
Then
\begin{align}\label{rank Z}
\Rank(\Zb_{n}^\star) &= \Rank \left( \Xb_n^{1/2}\Xb_n^{1/2} -
\frac{1}{\gamma_n}
    \begin{bmatrix}\Ib_{N_t} & \hhb_{nn}
    \end{bmatrix}\yb\yb^H
    \begin{bmatrix}\Ib_{N_t} \\\hhb_{nn}^H\end{bmatrix}\right)
    \notag \\
    &=\Rank\left(\Ib_{N_t}-\Xb_n^{-1/2}\frac{1}{\gamma_n}
    \begin{bmatrix}\Ib_{N_t} & \hhb_{nn}
    \end{bmatrix}\yb\yb^H
    \begin{bmatrix}\Ib_{N_t} \\\hhb_{nn}^H
    \end{bmatrix}\Xb_n^{-1/2}\right)
    \geq N_t-1.
\end{align}
It follows from \eqref{C1} and \eqref{rank Z} that
\begin{align}\label{rank W}
   0< \Rank(\Wb_{n}^\star) \leq N_t - \Rank(\Zb_n^\star) \leq 1,
\end{align} that is, $\Wb_{n}^\star$ must be of rank one.

What remains is to show that $\Yb_{nn}^\star$ is indeed of rank one. Firstly, one can show that $c_n >0$ since, if not, by \eqref{C9} and by the fact of $\Qb_{nn}\succ \zerob$, we must have $\Yb_{nn}^\star=\zerob$, which, however, will lead to $\Zb_{n}^\star\succ\zerob$ and $\Wb_n^\star=\zerob$, and thus contradicts with \eqref{C1}. By substituting \eqref{part Y} into \eqref{C3}, we obtain the following two equalities
\begin{align}
\left(\frac{1}{\gamma_n}\Wb_n^\star + \lambda_{nn}^\star\Qb_{nn}
\right)\Ab_n &+
\frac{1}{\gamma_n}\Wb_n^\star\hhb_{nn}\bb_n^H =\zerob,    \label{E0100-1}\\
\left({\frac{1}{\gamma_n}}\Wb_n^\star + \lambda_{nn}^\star\Qb_{nn}
\right)\bb_n &+ {\frac{1}{\gamma_n}}\Wb_n^\star\hhb_{nn}  c_n   =\zerob.
\label{E0100-2}
\end{align}
Further right multiplying \eqref{E0100-2} with $-\bb_n^H/c_n$, and adding the resultant {equality to} \eqref{E0100-1} gives rise to
\begin{align}
     \left(\frac{1}{\gamma_n}\Wb_n^\star + \lambda_{nn}^\star\Qb_{nn} \right)(\Ab_n-\bb_n\bb_n^H/c_n)=\zerob. \label{E0100-99}
\end{align}
Since $\left(\frac{1}{\gamma_n}\Wb_n^\star +
\lambda_{nn}^\star\Qb_{nn} \right) \succ \zerob$ due to both
$\lambda_{nn}^\star>0$ and $\Qb_{nn}\succ \zerob$, \eqref{E0100-99}
implies that $\Ab_n=\bb_n\bb_n^H/c_n$, and thus
\begin{align}\label{part Y2}
{\Yb}_{nn}^\star=\begin{bmatrix}\bb_n\bb_n^H/c_n&\bb_n\\\bb_n^H&c_n\end{bmatrix}=
\begin{bmatrix}\bb_n/\sqrt{c_n}\\ \sqrt{c_n}\end{bmatrix}[\bb_n^H/\sqrt{c_n}~
\sqrt{c_n}],
\end{align} which is a rank-one matrix. {Case C1) is thus} proved.

\emph{Proof of case C2):} Case C2) can be proved following similar derivations in \eqref{rank Z} and \eqref{rank W},
but using the KKT conditions of problem \eqref{RobustMCBF} with $\Qb_{nnk}=\infty\Ib_{N_t}$ (i.e., $\eb_{nnk}=\zerob$) for all $n,k$.

\emph{Proof of case C3):} Case C3) is a generalization of the result in \cite{Song2011} where the tightness of SDR for the worst-case robust beamforming problem in the single-cell scenario ($N_c=1$) is studied. {The condition in \eqref{conditions 1}} can be proved following exactly the same idea as in \cite{Song2011} and thus the details are omitted here.} \hfill $\blacksquare$

\vspace{-0.0cm}
\bibliography{RobustMCBFref}

\begin{thebibliography}{10}
\providecommand{\url}[1]{#1}
\csname url@samestyle\endcsname
\providecommand{\newblock}{\relax}
\providecommand{\bibinfo}[2]{#2}
\providecommand{\BIBentrySTDinterwordspacing}{\spaceskip=0pt\relax}
\providecommand{\BIBentryALTinterwordstretchfactor}{4}
\providecommand{\BIBentryALTinterwordspacing}{\spaceskip=\fontdimen2\font plus
\BIBentryALTinterwordstretchfactor\fontdimen3\font minus
  \fontdimen4\font\relax}
\providecommand{\BIBforeignlanguage}[2]{{%
\expandafter\ifx\csname l@#1\endcsname\relax
\typeout{** WARNING: IEEEtran.bst: No hyphenation pattern has been}%
\typeout{** loaded for the language `#1'. Using the pattern for}%
\typeout{** the default language instead.}%
\else
\language=\csname l@#1\endcsname
\fi
#2}}
\providecommand{\BIBdecl}{\relax}
\BIBdecl

\bibitem{Shen2011}
C.~Shen, K.-Y. Wang, T.-H. Chang, Z.~Qiu, and C.-Y. Chi, ``Worst-case {SINR}
  constrained robust coordinated beamforming for multicell wireless systems,''
  in \emph{Proc. IEEE ICC}, Kyoto, Japan, Jun. 5-9, 2011, pp. 1--5.

\bibitem{Gesbert10JSAC}
D.~Gesbert, S.~Hanly, H.~Huang, S.~S. Shitz, O.~Simeone, and W.~Yu,
  ``Multi-cell {MIMO} cooperative networks: A new look at interference,''
  \emph{IEEE J. Sel. Areas in Commun.}, vol.~28, no.~9, pp. 1380--1408, Dec.
  2010.

\bibitem{Irmer11COMM}
R.~Irmer, V.~H. Droste, D.~Telekom, P.~Marsch, M.~Grieger, and G.~Fettweis,
  ``Coordinate multipoint: {C}oncepts, performance, and field trial results,''
  \emph{IEEE Commun. Mag.}, vol.~49, no.~2, pp. 102--111, Feb. 2011.

\bibitem{Dahrouj10TWC}
H.~Dahrouj and W.~Yu, ``Coordinated beamforming for the multicell multi-antenna
  wireless system,'' \emph{IEEE Trans. Wireless Commun.}, vol.~9, no.~5, pp.
  1748--1759, May 2010.

\bibitem{Venturino10TWC}
L.~Venturino, N.~Prasad, and X.-D. Wang, ``Coordinated linear beamforming in
  downlink multi-cell wireless networks,'' \emph{IEEE Trans. Wireless Commun.},
  vol.~9, no.~4, pp. 1451--1461, Apr. 2010.

\bibitem{Bjornson2010}
E.~Bj{\"{o}}rnson, M.~Bengtsson, and B.~Ottersten, ``Optimality properties and
  low-complexity solutions to coordinated multicell transmission,'' in
  \emph{Proc. IEEE GLOBECOM}, Miami, FL, Dec. 6-10, 2010, pp. 1--6.

\bibitem{Nguyen2011TSP}
D.~H.~N. Nguyen and T.~Le-Ngoc, ``Multiuser downlink beamforming in multicell
  wireless systems: {A} game theoretical approach,'' \emph{IEEE Trans. Signal
  Process.}, vol.~57, no.~7, pp. 3326--3338, July 2011.

\bibitem{Tolli11TWC}
A.~T{\"{o}}lli, H.~Pennanen, and P.~Komulanen, ``Decentralized minimum power
  multi-cell beamforming with limited backhaul signaling,'' \emph{IEEE Trans.
  Wireless Commun.}, vol.~10, no.~2, pp. 570--580, Feb. 2011.

\bibitem{Zhangrui2010TSP}
R.~Zhang and S.~Cui, ``Cooperative interference management with {MISO}
  beamforming,'' \emph{IEEE Trans. Signal Process.}, vol.~58, no.~10, pp.
  5450--5458, Oct. 2010.

\bibitem{HuangTSP2011}
Y.~Huang, G.~Zheng, M.~Bengtsson, K.-K. Wong, L.~Yang, and B.~Ottersten,
  ``Distributed multicell beamforming with limited intercell coordination,''
  \emph{IEEE Trans. Signal Process.}, vol.~59, no.~2, pp. 728--738, Feb. 2011.

\bibitem{3GPP_standard}
Technical Specification Group Radio Access Network; High Speed Packet Access
  (HSPA) Evolution; Frequency Division Duplex {(FDD)}, {3GPP TR} 25.999,
  V7.1.0, Mar. 2008, available at
  http://www.3gpp.org/ftp/Specs/html-info/25999.htm.

\bibitem{Qiu2011TSP}
J.~Qiu, R.~Zhang, Z.-Q. Luo, and S.~Cui, ``Optimal distributed beamforming for
  {MISO} interference channels,'' in \emph{Proc. ASILOMAR}, Pacific Grove, CA,
  Nov. 7-10, 2010, pp. 277--281.

\bibitem{Farrokhi1998}
F.~Rashid-Farrokhi, K.~J.~R. Liu, and L.~Tassiulas, ``Transmit beamforming and
  power control for cellular wireless systems,'' \emph{IEEE J. Sel. Areas
  Commun.}, vol.~16, no.~8, pp. 1437--1450, Oct. 1999.

\bibitem{Boyddecomposition}
S.~Boyd, L.~Xiao, A.~Mutapcic, and J.~Mattingley, ``Notes on decomposition
  methods,'' avaliable at
  \url{http://see.stanford.edu/materials/lsocoee364b/08-decomposition_notes.pd%
f}.

\bibitem{Love08JSAC}
D.~Love, R.~Heath, V.~Lau, D.~Gesbert, B.~Rao, and M.~Andrews, ``An overview of
  limited feedback in wireless communication systems,'' \emph{IEEE J. Sel.
  Areas in Commun.}, vol.~26, no.~8, pp. 1341--1365, Oct. 2008.

\bibitem{Bjornson11TSP}
E.~Bj{\"{o}}rnson, G.~Zheng, M.~Bengtsson, and B.~Ottersten, ``Robust monotonic
  optimization framework for multicell {MISO} systems,'' \emph{submitted to
  IEEE Trans. Signal Process.}, avaliable at
  \url{http://arxiv.org/PS_cache/arxiv/pdf/1104/1104.5240v1.pdf}.

\bibitem{Tajer11TSP}
A.~Tajer and N.~P. adn X.-D.~Wang, ``Robust linear precoder design for
  multi-cell downlink transmission,'' \emph{IEEE Trans. Signal Process.},
  vol.~59, no.~1, pp. 235--251, Jan. 2011.

\bibitem{Shenouda07JSTSP}
M.~Shenouda and T.~Davidson, ``Convex conic formulations of robust downlink
  precoder designs with quality of service constraints,'' \emph{IEEE J. Sel.
  Topics Signal Process.}, vol.~1, no.~4, pp. 714--724, Dec. 2007.

\bibitem{ZhengWongNg_2008}
G.~Zheng, K.-K. Wong, and T.-S. Ng, ``Robust linear {MIMO} in the downlink: A
  worst-case optimization with ellipsoidal uncertainty regions,'' \emph{EURASIP
  J. Adv. Signal Process.}, vol. 2008, pp. 1--15, June 2008, {Article ID}
  609028.

\bibitem{Luo2010_SPM}
Z.-Q. Luo, W.-K. Ma, A.~M.-C. So, Y.~Ye, and S.~Zhang, ``Semidefinite
  relaxation of quadratic optimization problems,'' \emph{IEEE Signal Process.
  Mag.}, pp. 20--34, May 2010.

\bibitem{BK:LuoChang}
Z.-Q. Luo and T.-H. Chang, ``{SDP} relaxation of homogeneous quadratic
  optimization: {A}pproximation bounds and applications,'' in \emph{Convex
  Optimization in Signal Processing and Communications}, Chapter 4, UK:
  Cambridge University, 2010.

\bibitem{BK:BoydV04}
S.~Boyd and L.~Vandenberghe, \emph{Convex {O}ptimization}.\hskip 1em plus 0.5em
  minus 0.4em\relax Cambridge, UK: Cambridge University Press, 2004.

\bibitem{BertsekasADMM}
D.~P. Bertsekas and J.~N. Tsitsiklis, \emph{Parallel and distributed
  computation: {Numerical} methods}.\hskip 1em plus 0.5em minus 0.4em\relax
  Upper Saddle River, NJ, USA: Prentice-Hall, Inc., 1989.

\bibitem{BoydADMM}
S.~Boyd, N.~Parikh, E.~Chu, B.~Peleato, and J.~Eckstein, ``Distributed
  optimization and statistical learning via the alternating direction method of
  multipliers,'' \emph{Foundations and Trends in Machine Learning}, vol.~3,
  no.~1, pp. 1--122, 2008.

\bibitem{BK:Bertsekas06}
D.~P. Bertsekas and J.~N. Tsitsiklis, \emph{Nonlinear Programming: 2nd
  Edition}.\hskip 1em plus 0.5em minus 0.4em\relax Belmont, MA, USA: Athena
  Scientific, 1999.

\bibitem{Gershman2010_SPM}
A.~B. Gershman, N.~D. Sidiropoulos, S.~Shahbazpanahi, M.~Bengtsson, and
  B.~Ottersten, ``Convex optimization-based beamforming,'' \emph{IEEE Signal
  Process. Mag.}, pp. 62--75, May 2010.

\bibitem{Dahrouj2010}
H.~Dahrouj and W.~Yu, ``Coordinated beamforming for the multicell multi-antenna
  wireless system,'' \emph{IEEE Trans. Wireless Commun.}, vol.~9, no.~5, pp.
  1748--1759, May 2010.

\bibitem{SeDuMi}
J.~Sturm, ``{SeDuMi}: version 1.1,'' \url{http://sedumi.ie.lehigh.edu/}, Oct.
  2004.

\bibitem{Song2011}
E.~Song, Q.~Shi, M.~Sanjabi, R.~Sun, and Z.-Q. Luo, ``Robust {SINR}-constrained
  {MISO} downlink beamforming: {W}hen is semidefinite programming relaxation
  tight?'' in \emph{Proc. IEEE ICASSP}, Progue, Czech, May 22-27, 2011, pp.
  3096--2099.

\bibitem{Chang2011Asilomar}
T.-H. Chang, W.-K. Ma, and C.-Y. Chi, ``Worst-case robust multiuser transmit
  beamforming using semidefinite relaxation: {Duality} and implications,''
  submitted to Asilomar Conference on Signals, Systems, and Computers, Pacific
  Glove, CA, USA, Nov. 6-9, 2011.

\bibitem{Zhang08TWC}
H.~Zhang, N.~B. Mehta, A.~F. Molisch, J.~Zhang, and H.~Dai, ``Asynchronous
  interference mitigation in cooperative base station systems,'' \emph{IEEE
  Trans. Wireless Commun.}, vol.~7, no.~1, pp. 155--165, Jan. 2008.

\bibitem{3GPP_standard2}
Technical Specification Group Radio Access Network; Physical layer aspects for
  evolved Universal Terrestrial Radio Access (UTRA), 3GPP TR 25.814, v.7.1.0,
  Sep. 2006, available at http://www.3gpp.org/ftp/Specs/html-info/25814.htm.

\bibitem{Huh2010}
H.~Huh, H.~C. Papadopoulos, and G.~Caire, ``Multiuser {MISO} transmitter
  optimization for intercell interference mitigation,'' \emph{IEEE Trans.
  Signal Process.}, vol.~58, no.~8, pp. 4272 --4285, Aug. 2010.

\end{thebibliography}
\end{document}